\documentclass[11pt]{article}
\pdfoutput=1 
\usepackage{amsmath,amssymb,amsbsy,amstext, amsthm, simplewick, amsfonts}
\usepackage{graphicx}
\usepackage{epsfig}
\usepackage{verbatim} 
\usepackage{fancybox}
\usepackage{pgfplots}
\pgfplotsset{width=10cm,compat=1.9}
\usepackage{color}
\usepackage{mathdots}
\usepackage{ulem}
\usepackage{slashed}
\usepackage{enumitem}
\usepackage{caption}
\usepackage{subfig}
\usepackage{mathtools}
\usepackage{youngtab}
\usepackage{bbm}
\usepackage{parskip}
\usepackage{booktabs}
\usepackage[numbers,sort&compress]{natbib}
\usepackage{youngtab}
\usepackage{ytableau}
\usepackage{genyoungtabtikz}
\usepackage[utf8]{inputenc}
\usepackage{tikz}
\usetikzlibrary{positioning, trees,decorations, decorations.markings, decorations.pathmorphing, arrows, shapes.geometric}

\allowdisplaybreaks

\newcommand{\Comment}[1]{{}}
\definecolor{darkblue}{rgb}{0.15,0.35,0.55}
\definecolor{reddish}{rgb}{0.65, 0.2, 0.2}
\usepackage[linktocpage=true]{hyperref}
\hypersetup{
colorlinks=true,
citecolor=darkblue,
linkcolor=reddish,
urlcolor=darkblue,
pdfauthor={},
pdftitle={},
pdfsubject={}
}

\usepackage{cleveref}

\newcommand{\be}{\begin{equation}}
\newcommand{\ee}{\end{equation}}
\newcommand{\bea}{\begin{eqnarray}}
\newcommand{\eea}{\end{eqnarray}}
\newcommand{\nn}{\nonumber}

\newcommand*\circled[1]{\tikz[baseline=(char.base)]{
            \node[shape=circle,draw,inner sep=.8pt] (char) {  #1};}}

\numberwithin{equation}{section}

\makeatletter
\def\thickhline{%
  \noalign{\ifnum0=`}\fi\hrule \@height \thickarrayrulewidth \futurelet
   \reserved@a\@xthickhline}
\def\@xthickhline{\ifx\reserved@a\thickhline
               \vskip\doublerulesep
               \vskip-\thickarrayrulewidth
             \fi
      \ifnum0=`{\fi}}
\makeatother

\newlength{\thickarrayrulewidth}
\begin{document}

\renewcommand{\thefootnote}{\fnsymbol{footnote}}
~
\vspace{0truecm}
\thispagestyle{empty}
\begin{center}
{\fontsize{19}{18} \bf{Fermionic Shift Symmetries in (Anti) de Sitter Space}}
\end{center} 

\vspace{.7truecm}

\begin{center}
{\fontsize{13.5}{18}\selectfont
James Bonifacio${}^{\rm a,}$$^{}$\footnote{E-mail: \Comment{\href{mailto:bonifacio@phy.olemiss.edu}}{\tt bonifacio@phy.olemiss.edu}}  and Kurt Hinterbichler${}^{\rm b,}$$^{}$\footnote{E-mail: \Comment{\href{mailto:kurt.hinterbichler@case.edu}}{\tt kurt.hinterbichler@case.edu}} 
}
\end{center}
\vspace{.4truecm}

 \centerline{{\it ${}^{\rm a}$Department of Physics and Astronomy,}}
 \centerline{{\it University of Mississippi, University, MS 38677}} 
 
   \vspace{.3cm}

 \centerline{{\it ${}^{\rm b}$CERCA, Department of Physics,}}
 \centerline{{\it Case Western Reserve University, 10900 Euclid Ave, Cleveland, OH 44106}} 
 
  \vspace{.3cm}

\vspace{.3cm}
\begin{abstract}
\noindent

We study extended shift symmetries that arise for fermionic fields on anti-de Sitter (AdS) space and de Sitter (dS) space for particular values of the mass relative to the curvature scale.  We classify these symmetries for general mixed-symmetry fermionic fields in arbitrary dimension and describe how fields with these symmetries arise as the decoupled longitudinal modes of massive fermions as they approach partially massless points.  For the particular case of AdS$_4$, we look for non-trivial Lie superalgebras that can underly interacting theories that involve these fields.  We study from this perspective the minimal such theory, the Akulov--Volkov theory on AdS$_4$, which is a non-linear theory of a spin-$1/2$ Goldstino field that describes the spontaneous breaking of ${\cal N}=1$ supersymmetry on  AdS$_4$ down to the isometries of  AdS$_4$. We show how to write the nonlinear supersymmetry transformation for this theory using the fermionic ambient space formalism.
We also study the Lie superalgebras of candidate multi-field examples and rule out the existence of a supersymmetric special galileon on AdS$_4$.

\end{abstract}

\newpage

\setcounter{tocdepth}{2}
\tableofcontents
\newpage

\renewcommand*{\thefootnote}{\arabic{footnote}}
\setcounter{footnote}{0}

\section{Introduction}

Shift symmetries are a hallmark of theories with spontaneously broken symmetries.  Spontaneously broken symmetries give rise to Goldstone bosons, and under the action of the broken generators the Goldstone fields transform by simple shifts, at leading order in powers of the field.  
On de Sitter (dS) and anti-de Sitter (AdS) space, in the case where the symmetry breaking involves spacetime symmetries, there is a class of allowed shift symmetries compatible with the underlying (A)dS isometries.  These shift symmetries generalize the extended shift symmetries of the flat space galileons \cite{Luty:2003vm,Nicolis:2008in} to (A)dS space and to higher spins.  The fields possessing these symmetries can be thought of as the longitudinal modes of massive fields that decouple as various massless and partially massless limits are taken in (A)dS.

On flat space, Goldstone fields and their generalizations with extended shift symmetries are massless.  A feature of the (A)dS shift-symmetric fields is that they have masses; the shift symmetries appear only at particular discrete values of the masses relative to the background curvature, with different mass values corresponding to different types of shift symmetries.  We can classify the different shift symmetries by a non-negative integer $k$, which gives the number of factors of the spacetime coordinates that appear in the transformation law, e.g., for flat space scalars $k=0$ is a simple constant shift, $k=1$ is the galileon, and $k=2$ is the special galileon \cite{Cheung:2014dqa,Hinterbichler:2015pqa,Cheung:2016drk,Novotny:2016jkh}.
Specific examples of shift symmetries of this kind on (A)dS have been discussed for many years in widely varying contexts \cite{vilenkin1978special,Allen:1985ux,Allen:1987tz,Antoniadis:1991fa,Folacci:1992xc,Folacci:1996dv,Shaynkman:2000ts,Gazeau:2010mn,Bros:2010wa,Goon:2011qf,Goon:2011uw,Burrage:2011bt,Epstein:2014jaa,Chekmenev:2015kzf,deBoer:2015kda, deBoer:2016pqk,Baumann:2017jvh,Arkani-Hamed:2018kmz}.
The possible shift symmetries for all symmetric tensor bosons were classified in Ref.~\cite{Bonifacio:2018zex} and this was extended to bosonic anti-symmetric and mixed-symmetry fields in Ref.~\cite{Hinterbichler:2022vcc}.   

In this paper, we extend this classification to fermionic fields.  Fermionic shift symmetries are associated with broken symmetry generators that are fermionic, so this is the case of broken supersymmetry (SUSY) and the fermions which transform with shifts under the broken generators are Goldstinos.  The canonical example of such a theory is the Akulov--Volkov theory \cite{Volkov:1972jx,Volkov:1973ix}. On flat space, this theory describes the low-energy dynamics of the spin-1/2 Goldstino mode resulting from the spontaneous breaking of ${\cal N}=1$ SUSY down to Poincar\'e symmetry. The Goldstino is massless and to leading order in powers of the fields it transforms with a simple shift under the broken SUSY generator.

On (A)dS space, the Akulov--Volkov Goldstino acquires a specific mass of order the background curvature scale \cite{Deser:1977uq, Zumino:1977av}.  This is the simplest example of the general classification of shift symmetries acting on fermionic fields on (A)dS that we describe.  As with bosonic fields, these shift symmetries occur for all possible fermionic spins, and they appear at particular discrete values of the fermion masses relative to the background curvature.  The shift-symmetric fields can be thought of as the longitudinal modes of massive fermionic fields that decouple as the masses approach massless or partially massless points.  From this perspective, the Akulov--Volkov Goldstino is the longitudinal mode of a massive spin-3/2 field that decouples as the mass approaches the massless value.  Indeed, in the case of spontaneously broken gauged SUSY, i.e., supergravity (SUGRA), this is realized physically as the super-Higgs mechanism: the Goldstino is eaten by the massless spin-3/2 field in the SUGRA gravity multiplet to become a massive spin-3/2 field, which uses the degrees of the freedom of the Goldstino for its longitudinal modes \cite{Deser:1977uq,Casalbuoni:1988sx}.

On AdS space, the fermionic shift-symmetric fields are unitary---they have half-integer dual conformal dimensions which lie above the relevant unitarity bound.  On dS space, they are non-unitary---they do not appear in the classification of fermionic dS unitary representations \cite{10.1063/1.1665471}.

The above classification of possible (A)dS shift symmetries is at the level of free field theory or representation theory.  A more interesting, and more difficult, dynamical question is whether there exist non-trivial interactions that preserve the shift symmetries.  
The free shift-symmetric fields always have shift-invariant field strengths constructed from derivatives of the field.
Trivial interactions can always be formed by using simple powers of the shift-invariant field strength, analogous to Euler--Heisenberg or Born--Infeld type interactions in $U(1)$ gauge theory.  These interactions retain the same shift symmetry as the linear theory.  By `non-trivial' interactions, we mean interactions that are not of this type, but rather those that deform the symmetry algebra so that the shift-symmetries do not commute.  The $k=1$ and $k=2$ scalars allow non-trivial self-interactions of this type
 \cite{Goon:2011qf,Goon:2011uw,Bonifacio:2018zex,Bonifacio:2021mrf}, as does the $k=0$ vector \cite{DeRham:2018axr,Bonifacio:2019hrj}, but no other examples are currently known for the bosonic theories.  
 
 A necessary condition for non-trivial interactions in the above mentioned sense is the existence of a suitable Lie algebra that underlies the symmetries \cite{Bogers:2018zeg,Roest:2019oiw}.  Trivial interactions do not deform the algebra of the linear theory, whereas non-trivial interactions do deform the algebra.  The interacting examples mentioned above are single field cases whose algebra allows for a deformation in which the shift-symmetries do not commute \cite{Bonifacio:2018zex}.
 
 Below we study the question of whether non-trivial interactions can exist for shift-symmetric fermions.  We do this by studying whether there exist suitable Lie superalgebras that deform the Lie superalgebra of the free theories.
 The properties of these Lie superalgebras are dimension and signature dependent, so for clarity we will stick to the specific case of four-dimensional AdS space.  AdS$_4$ has the isometry algebra $\mathfrak{so}(2,3)$, which is isomorphic to the symplectic Lie algebra $\mathfrak{sp}(4, \mathbb{R})$.

 By studying these algebras, we can rule out the existence of certain theories.  One example is the SUSY special galileon on AdS space.
The possibility of supersymmetric galileons in flat spacetime has been much studied \cite{Khoury:2011da,Koehn:2013hk,Farakos:2013fne,Kamimura:2013mia,Queiruga:2016yzd,Elvang:2017mdq,Roest:2017uga,Deen:2017dpm,Deen:2017jqv}. 
The special galileon in AdS$_4$ is a scalar field theory with a $k=2$ shift transformation that realizes the symmetry breaking pattern $\mathfrak{sl}(5, \mathbb{R}) \rightarrow \mathfrak{so}(3,2)$  \cite{Bonifacio:2018zex, Bonifacio:2021mrf}. It is known that there is no supersymmetric special galileon in flat spacetime \cite{Elvang:2018dco,Roest:2019dxy}.  Here, by studying the possible Lie superalgebras, we will see that there is no suitable symmetry algebra that could underly a non-trivial SUSY special galileon on AdS$_4$, and so no such theory exists.  
 
 The simplest suitable Lie superalgebra that does exist has a single fermionic shift symmetry acting on a spin-1/2 fermion, in addition to the AdS isometries.  This Lie superalgebra corresponds to $\mathfrak{osp}(1 | 4)$, the algebra of $\mathcal{N}=1$ supersymmetry in AdS$_4$.  This is the algebra underlying the AdS Akulov--Volkov theory.
 We will discuss the symmetry transformations of this theory in detail using the AdS$_4$ fermionic ambient space formalism.
 
 Another such Lie superalgebra is $\mathfrak{su}(2,2|1)$, which has two spin-1/2 fermionic generators, a vector generator and two scalar generators, in addition to the generators of the AdS isometries.  
This has an $\mathfrak{osp}(1 | 4)$ subalgebra which, when linearly realized, corresponds to the breaking pattern $\mathfrak{su}(2,2|1) \rightarrow \mathfrak{osp}(1|4)$, whose bosonic part is $\mathfrak{so}(4,2) \times \mathfrak{u}(1) \rightarrow \mathfrak{so}(3,2)$.  This underlies the AdS$_4$ supersymmetric DBI theory \cite{Ivanov:1979ft}, in which five-dimensional AdS SUSY is broken to four-dimensional AdS SUSY.  
This is a multi-field example: there is a $k=1$ shift-symmetric scalar (the DBI scalar), a $k=0$ scalar, and a $k=0$ shift-symmetric spin-1/2 fermion.  These three fields together form a linear shift-symmetric multiplet of ${\cal N}=1$ AdS$_4$ SUSY.  These fields can also be thought of as the longitudinal modes that decouple when an ${\cal N}=1$ massive graviton multiplet approaches the limit of the partially massless multiplet  \cite{Garcia-Saenz:2018wnw,Buchbinder:2019olk,Bittermann:2020xkl}.

Once multiple fields are considered, there is an infinite series of candidate Lie superalgebras that can underlie theories with multiple shift-symmetric fields including fermions.  These can be constructed by considering the associative superalgebra of endomorphisms of a $\mathbb{Z}_2$-graded vector space and using the natural (anti)commutator to form the Lie superbracket.
 These are analogous to the bosonic algebras studied in Ref.~\cite{Joung:2015jza}, which are finite-dimensional subalgebras of the infinite-dimensional higher-spin algebras of partially massless Vasiliev theories \cite{Bekaert:2013zya,Basile:2014wua,Alkalaev:2014nsa,Joung:2015jza,Brust:2016zns}.  The existence of these algebras allows for the existence of the corresponding non-linear shift-symmetric theories, but we do not have a practical method to construct the theories explicitly, so we cannot say for certain whether they exist. 

The rest of this paper is structured as follows: in Section~\ref{sec:free-shifts}, we describe the fermionic shift-symmetric representations for (A)dS space in general dimensions, including mixed-symmetry fields. In Section~\ref{sec:ambient}, we describe the fermionic ambient space formalism for AdS$_4$, making use of the bispinor formalism from Ref.~\cite{Binder:2020raz}. We then use this formalism to write the free shift symmetry transformations for AdS$_4$ in Section~\ref{sec:AdS4-shifts}.  In Section~\ref{sec:superalgebras}, we discuss Lie superalgebras deforming the symmetry algebras of the free theories, and we use the ambient space formalism to write symmetry transformations for the Akulov--Volkov theory in Section~\ref{AVsection}. We conclude in Section~\ref{sec:conclusions}.

\section{Shift symmetries of free fermions}
\label{sec:free-shifts}

In this section, we discuss the generalities and classification of shift-symmetric free fermions on (A)dS in arbitrary spacetime dimension $D$.  In order to keep the dimension general, we will stick to Dirac fermions, bearing in mind that these can be subject to further reducibility conditions (Majorana, Weyl, etc.) depending on the value of $D$ mod 8.  In the later sections, in order to explicitly write the symmetry transformations and discuss the algebras, we will specialize to AdS$_4$ using the bispinor formalism from Ref.~\cite{Binder:2020raz}. 

Consider $D$-dimensional anti-de Sitter space AdS$_D$ of radius $L$, where $D \geq 3$.  The Ricci scalar is $R=-D(D-1)L^{-2}$, the spacetime indices $\mu$ take values $0, \dots D-1$, and the metric is $g_{\mu\nu}$.  The gamma matrices $\gamma^{\mu}$ are the curved space $D$-dimensional gamma matrices satisfying
\be 
\left\{\gamma_\mu,\gamma_\nu\right\}=2g_{\mu\nu}.\label{gammalebgeraee}
\ee
They are related to the flat gamma matrices $\gamma^a$ satisfying $\left\{\gamma_a,\gamma_b\right\}=2\eta_{ab}$ by contracting with the AdS$_D$ vielbein $e_\mu^{\ a}$, $\gamma_\mu=e_\mu^{\ a}\gamma_a$.

The fermions will be described by tensor-spinors $\Psi_{\mu_1\mu_2\cdots}$ which carry some spacetime indices in addition to a single Dirac fermion index, where the fermion index will generally be suppressed in this section.
The covariant derivative ${\cal D}_\mu$ operating on a tensor-spinor includes the connection for the spinor, i.e., ${\cal D}_\mu=\nabla_\mu+{1\over 8}\omega_\mu^{ab}\left[\gamma_a,\gamma_b\right]$, where $\omega_\mu^{ab}$ is the Lorentz spin connection and $\nabla_\mu$ is the standard tensor covariant derivative containing the Christoffel symbols acting on the tensor indices.  The covariant derivative ${\cal D}_\mu$ commutes with $\gamma^\mu$ and with $g_{\mu\nu}$.  We also define 
\be 
\slashed{\cal D}\coloneqq \gamma^\mu {\cal D}_\mu
\ee
and the following (A)dS modified derivative, which appears often:
\be 
\tilde{\cal D}_\mu={\cal D}_\mu\pm {1\over 2L}\gamma_\mu.\label{dtildedefcee}
\ee
(The sign ambiguity will correspond to a choice of sign for partially massless and shift-symmetric fermion mass terms.)

For a general tensor-spinor $\Psi_{\mu_1\cdots \mu_{r}}$ of rank $r$, we have the following useful identities:
\be  \left[ {\cal D}_\mu,{\cal D}_\nu\right]\Psi_{\mu_1\cdots \mu_{r}}=-{1\over 2L^2}\gamma_{\mu\nu}\Psi_{\mu_1\cdots \mu_r}-  {2\over L^2} \sum_{i=1}^{r} g_{\mu_i[\mu}\Psi_{| \mu_1\cdots \mu_{i-1} | \nu] \mu_{i+1} \cdots \mu_{r}} \, ,\ee
\be  \slashed{\cal D}^2\Psi_{\mu_1\cdots \mu_{r}}=\left({\cal D}^2+{1\over L^2}\left[r+{D(D-1)\over 4}\right]\right)\Psi_{\mu_1\cdots \mu_{r}}-{1\over L^2} \sum_{i=1}^{r} \gamma_{\mu_i}\gamma^{\mu}\Psi_{ \mu_1\cdots \mu_{i-1}  \mu \mu_{i+1} \cdots \mu_{r}}  \, ,\label{D2identityee}
\ee
where we antisymmetrize with weight one, e.g., $T_{[\mu \nu]} = \frac{1}{2}(T_{\mu \nu} - T_{\nu \mu})$, and $\gamma_{\mu \nu}\equiv \gamma_{[\mu}\gamma_{\nu]}$.

\subsection{Spin-$s$ massive fermions}

A massive spin-$s$ fermion is described by a rank-$(s-\frac{1}{2})$ tensor-spinor field $\Psi_{\mu_1\cdots \mu_{s-1/2}}$.  The $\mu$'s are fully symmetric spacetime tensor indices and the single spinor index is suppressed. 
On shell, this field is transverse, gamma traceless, and satisfies the Dirac equation with a mass parameter that we call $\tilde m$,
\be  \label{eq:fermion-EOM}
\left(\slashed{\cal D}+\tilde m \right)\Psi_{\mu_1\cdots \mu_{s-1/2}}=0,\ \ \ {\cal D}^{\mu_1} \Psi_{\mu_1\cdots \mu_{s-1/2}}=0,\ \ \ \gamma^{\mu_1}\Psi_{\mu_1\cdots \mu_{s-1/2}}=0\,. 
\ee

Multiplying the gamma tracelessness condition by another gamma matrix and using \eqref{gammalebgeraee} shows that $\Psi_{\mu_1\cdots \mu_{s-1/2}}$ is also completely traceless in all its tensor indices, 
\be g^{\mu_1\mu_2}\Psi_{\mu_1\cdots \mu_{s-1/2}}=0.\ \ \ee
Applying $\slashed{\cal D}-\tilde m$ to the Dirac equation and using \eqref{D2identityee} along with gamma tracelessness, we see that the tensor-spinor also satisfies 
the Klein--Gordon equation,
\be \left({\cal D}^2 +{1\over L^2}\left[s-{1\over 2}+{D(D-1)\over 4}\right]-\tilde m^2\right) \Psi_{\mu_1\cdots \mu_{s-1/2}}=0\,.\ee

The AdS/CFT formula relating $\tilde m$ and the scaling dimension $\Delta$ of the dual CFT operator is
 \be \tilde m^2 L^2 =\left(\Delta-{d\over 2 }\right)^2\,,\ \ \Delta_\pm= {d\over 2}\pm \tilde m L\,,\ee
where $d =D-1$ is the boundary dimension. Note that this relation is independent of the spin.

\subsection{Spin-$s$ partially massless fermions}

Fermions acquire partially massless (PM) fermionic gauge symmetries for certain special masses when $s \geq \frac{3}{2}$ \cite{Deser:2001pe,Deser:2001xr,Deser:2003gw}. These PM values occur for the mass values
\be 
{\tilde m}_{s,t}^2 =  \frac{\left(D-3+2t\right)^2}{4L^2}, \ \ \ \ \label{PMfmassvaluese}
\ee
where 
\be t=\frac{1}{2},\frac{3}{2},\cdots,s-1\ee
is called the depth and labels the spin of the PM gauge parameter. A depth-$t$ spin-$s$ PM field has an on-shell gauge symmetry
\be \label{eq:PMsymmetry}
\delta \Psi_{\mu_1\cdots \mu_{s-\frac{1}{2}}} = \tilde{\cal D}_{(\mu_{t+\frac{1}{2}}} \dots \tilde {\cal D}_{\mu_{s-\frac{1}{2}}} \xi_{\mu_1 \dots \mu_{t-\frac{1}{2}})^T},
\ee
where $\xi_{\mu_1 \dots \mu_{t-\frac{1}{2}}}$ is a symmetric tensor-spinor spin-$t$ PM gauge parameter and $\tilde{\cal D}_\mu$ is the modified covariant derivative defined in \eqref{dtildedefcee}.  The gauge parameter itself satisfies on-shell conditions analogous to \eqref{eq:fermion-EOM},
\be \left(\slashed{\cal D}+\tilde m_\xi \right) \xi_{\mu_1 \dots \mu_{t-\frac{1}{2}}}=0,\ \ \ {\cal D}^{\mu_1}\xi_{\mu_1 \dots \mu_{t-\frac{1}{2}}}=0,\ \ \ \ \  \gamma^{\mu_1}\xi_{\mu_1 \dots \mu_{t-\frac{1}{2}}}=0\,, \label{gaugepree} \ee 
with 
\be \tilde m_\xi^2={\left(D-3+2s\right)^2\over 4L^2}\,.\label{ximassfdefe} \ee

The PM mass values \eqref{PMfmassvaluese} and the mass \eqref{ximassfdefe} of the Dirac equation in \eqref{gaugepree} are determined by the PM gauge symmetry \eqref{eq:PMsymmetry} up to an overall sign ambiguity: this sign corresponds to the choice of sign made in the definition of $\tilde {\cal D}_\mu$ in \eqref{dtildedefcee} used in \eqref{eq:PMsymmetry}, with the upper sign corresponding to taking the positive roots of \eqref{PMfmassvaluese} and \eqref{ximassfdefe}. 

The conformal dimensions corresponding to the PM mass values are
\be
 {\Delta}_{+}= d+t-1, \quad {\Delta}_{-}= 1-t\,.
\ee

The fields with the largest gauge symmetry, and hence the smallest number of propagating degrees of freedom, are those with $t=s-1$, which are the fields typically called ``massless'' \cite{Fang:1979hq,Aragone:1980rk,Vasiliev:1987tk}. The mass parameter $\tilde m$ in the Dirac equation is related to the traditionally defined mass $m$ by 
\be 
\tilde m^2=\frac{(D+2 s-5)^2}{4 L^2}+m^2.\label{massmtrelatione}
\ee
The mass $m$ is defined so that a  spin-$s$ field with $s\geq 3/2$ and with $m=0$ corresponds to the mass value with the largest gauge symmetry (for $s=1/2$, we do not have this natural definition of masslessness, but we can still use \eqref{massmtrelatione} as a definition in this case).  
The AdS/CFT formula in terms of $m^2$ is 
\be 
m^2L^2=(\Delta+s-2)(\Delta-s-d+2)\,,\ \ \ \Delta_\pm=   {d\over 2}\pm\sqrt{{\left(d+2(s-2)\right)^2\over 4}+m^2L^2}\,, \label{dimeqo}
\ee
and the PM mass values are
\be { m}_{s,t}^2 = \frac{(t-s+1) (D+s+t-4)}{L^2}\,.\ee
Note that the PM values for $m^2$ are negative except for the massless case, i.e., ${m}_{s,t}^2<0$ for $t<s-1$.

There is a unitarity bound for CFT primaries which for symmetric tensor-spinor fermions of the type we are considering is
\be \Delta \geq d+s-2.\ee
The massless fields saturate this bound and all the other PM points lie below this bound.  Thus on AdS space, the massless point is unitary and all the other PM points are non-unitary.  On dS space, where $L^2\rightarrow-1/H^2$ with $H$ the Hubble scale, the question of unitarity is more complicated and involves subtleties in the fermionic representation theory of the dS group \cite{10.1063/1.1665471,Basile:2016aen}.  It turns out that in $D=4$, the massless and PM points are all unitary (despite the fact that the Lagrangian is seemingly non-Hermitian due to the imaginary mass), but for $D\geq 5$ none of the massless or PM points are unitary (see, e.g., the recent study \cite{Letsios:2023qzq}). 

\subsection{Spin-$s$ shift-symmetric fermions}

A spin-$s$ fermion field acquires an enhanced shift symmetry of level $k$ at the mass value
\be 
\tilde{\mathsf{m}}^2_{s,k}= { \left(2k+2s+D-1\right)^2\over 4L^2},\ \ \  \mathsf{m}^2_{s, k} = {(k+2)(k+D-3+2s)\over L^{2}}. \label{shiftfermsymtmasse}
\ee
The level $k=0,1,2, \dots$ of the shift symmetry corresponds to the power of spacetime coordinates appearing in the shift symmetry.  The form of these shift symmetries is most naturally written in ambient space: the fermion shifts by a fermionic ambient space tensor-spinor whose tensor indices are in the finite-dimensional representation of the (A)dS isometry group corresponding to a traceless two-row Young tableau $[s-{1\over 2}+k,s-{1\over 2}]_{\rm s}$.\footnote{Young tableaux are denoted by $[s_1,s_2,\cdots,s_p]$, where $s_i$ is the number of boxes in the $i$-th row.  In this section, we include a subscript ${\rm s}$, $[s_1,s_2,\cdots,s_p]_{\rm s}$, to indicate that we are talking about the symmetries of a tensor-spinor, where the tensor indices are in the given tableau.} The indices of the top row are contracted with ambient space coordinates $X^A$ and the rest are projected down to the  (A)dS space.  The details of this projection and the irreducibility conditions that can be imposed on the spinor are dimension dependent, so we will not attempt to do this explicitly for general dimension.  We consider the case of AdS$_4$ in Section~\ref{sec:AdS4-shifts}.  These objects can be thought of as (A)dS spinor spherical harmonics, or conformal Killing tensor-spinors \cite{Campoleoni:2017vds}. 

The shift-symmetric scaling dimensions corresponding to the masses \eqref{shiftfermsymtmasse} are
\be 
\Delta_+ = d+s+k,\quad \Delta_- = -s-k.
\ee
 Similarly to the bosonic case, the negative root $\Delta_-=-s-k$ of the shift-symmetric fermion gives the representation whose non-null states span the finite-dimensional, non-unitary tensor-spinor representation $[s-{1\over 2}+k,s-{1\over 2}]_{\rm s}$ of the AdS$_D$ isometry group.
On AdS space, the shift-symmetric points all lie above the unitarity bound and hence are unitary.  On dS space, their absence from the list of unitary representations \cite{10.1063/1.1665471} indicates that they are non-unitary.

\begin{figure}[ht!]
\begin{center}
\begin{tikzpicture}
\node at (9.6, 0.52){$t=0$};
\draw [thick, dotted] (8.3,0.51) -- (9.04 ,0.51);
\node at (9.6, 1.01){$t=\frac{1}{2}$};
\draw [thick, dotted] (8.3,1.02) -- (9.04 ,1.02);
\node at (9.6, 1.54){$t=1$};
\draw [thick, dotted] (8.3,1.53) -- (9.04 ,1.53);
 \node[rotate=45] at (6.6,7.6) {$k=0$};
\draw [thick, dotted] (5.88,6.83) -- (6.25,7.21);
 \node[rotate=45] at (5.4,7.6) {$k=1$};
\draw [thick, dotted] (4.68,6.83) -- (5.05, 7.21);
 \node[rotate=45] at (4.29,7.6) {$k=2$};
\draw [thick, dotted] (3.57,6.83) -- (3.94, 7.21);
 \node[rotate=45] at (8.85,7.65) {{\small massless}};
\draw [thick, dotted] (8.13,6.83) -- (8.4,7.11);
	\begin{axis}[%
	axis y line=middle, axis x line=bottom,
	xlabel=$s$, xlabel near ticks,
	ylabel= $\Delta$, ylabel near ticks,
	xmin=0, xmax = 7.4, ymax=7.4, ymin=1/2,
	yticklabels={$d-1$, $d$, $d+1$, $d+2$, $d+3$, $d+4$, $d+5$},ytick={1,...,7},
	xtick={0, ...,7},
	scatter/classes={%
		a={mark=*,blue},%
		b={mark=o, thick, blue},%
		c={mark=o, thick, draw=black}}]
\addplot[black, thick, dotted, domain=0:6] {x+1};
	\addplot[scatter,only marks,%
		scatter src=explicit symbolic]%
	table[meta=label] {
x     y      label
1   1   a 
2   1   a 
3   1   a 
4   1   a 
5   1   a 
6   1   a 
7   1   a 
2   2   a 
3   2   a 
4   2   a 
5   2   a 
6   2   a 
7   2   a 
3   3   a 
4   3   a 
5   3   a 
6   3   a 
7   3   a 
4   4   a 
5   4   a 
6   4   a 
7   4   a 
5   5   a 
6   5   a 
7   5   a 
6   6   a 
7   6   a 
7 7 a
1.5 1.5   b
2.5 1.5   b
3.5 1.5   b
4.5 1.5   b
5.5 1.5   b
6.5 1.5   b
2.5 2.5   b
3.5 2.5   b
4.5 2.5   b
5.5 2.5   b
6.5 2.5   b
3.5 3.5   b
4.5 3.5   b
5.5 3.5   b
6.5 3.5   b
4.5 4.5   b
5.5 4.5   b
6.5 4.5   b
5.5 5.5   b
6.5 5.5   b
6.5 6.5   b
0 2 d
0 3 d
0 4 d
0 5 d
0 6 d
0 7 d
1 3 d
1 4 d
1 5 d
1 6 d
1 7 d
2 4 d
2 5 d
2 6 d
2 7 d
3 5 d
3 6 d
3 7 d
4 6 d
4 7 d
5 7 d
.5 2.5 c
.5 3.5 c
.5 4.5 c
.5 5.5 c
.5 6.5 c
1.5 3.5 c
1.5 4.5 c
1.5 5.5 c
1.5 6.5 c
2.5 4.5 c
2.5 5.5 c 
2.5 6.5 c
3.5 5.5 c
3.5 6.5 c
4.5 6.5 c
	};
	\end{axis}
\end{tikzpicture}
\caption{Scaling dimensions of PM fields and shift-symmetric fields on AdS$_{d+1}$. Filled circles are bosons and unfilled circles are fermions.  The blue circles below the dotted diagonal line are PM fields and the black circles above it are shift-symmetric fields.  The shift-symmetric field obtained in the decoupling limit as one approaches a PM field is obtained by reflecting the PM field about the dotted diagonal line $\Delta=s+d-1$. }
\label{fig:fermion-deltas}
\end{center}
\end{figure}
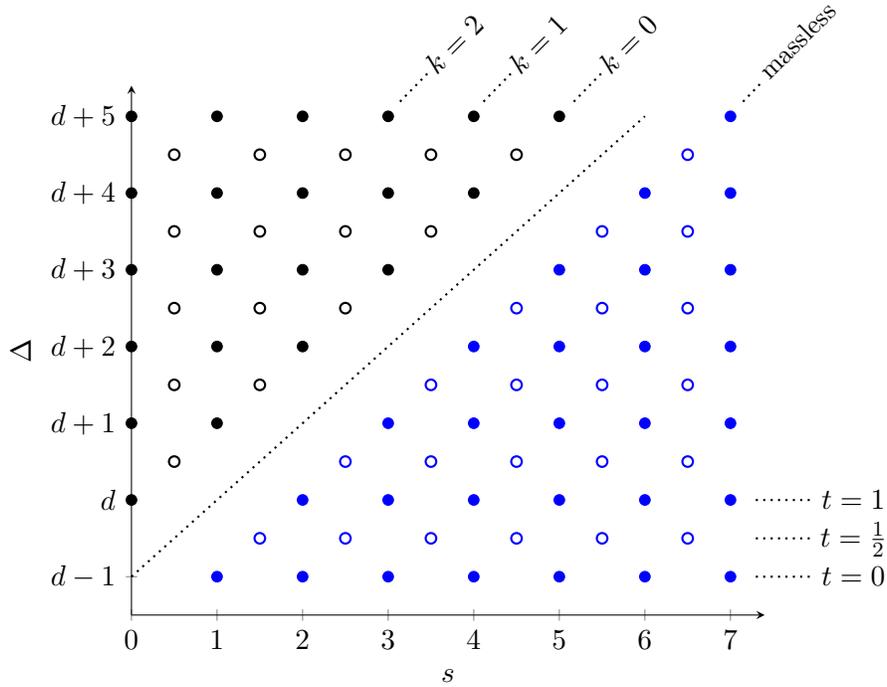

The connection between the fermionic PM fields and the fermionic shift-symmetric fields is analogous to that of the bosons: the shift-symmetric fields appear as the longitudinal modes of massive fields that decouple in the partially massless limits.  Consider taking the depth-$t$ PM limit of a fermionic spin-$s$ massive field. In this limit, the massive fermion breaks up into a depth-$t$ PM fermion and a spin-$t$ fermion with a level $k=s-t-1$ shift symmetry,
\be
(m^2, s) \xrightarrow[m^2 \rightarrow {m}^{2}_{s,t}]{} (m^2_{ s,t}, s)  \oplus ( \mathsf{m}^2_{ t, s-t-1}, t).
\ee
This corresponds to the following branching rule:
\be D (\Delta,s) \xrightarrow[\Delta\rightarrow d-1+t]{} D (d-1+t,s)\oplus D (s+d-1,t)\,.\ee 
From this viewpoint, the shift symmetries originate as the reducibility parameters of the PM gauge theory, i.e., those gauge parameters for which the PM transformation \eqref{eq:PMsymmetry} vanishes.  
We illustrate the scaling dimensions of the PM and shift-invariant fermions and bosons in Figure~\ref{fig:fermion-deltas}.

There is a shift-invariant field strength tensor that is patterned after the partially massless gauge symmetry \eqref{eq:PMsymmetry} of the parent PM field. For spin-$s$ and level $k$, we have
\be  F_{\mu_1\cdots \mu_{s-\frac{1}{2}+k+1}} = \tilde{\cal D}_{(\mu_{s-\frac{1}{2}+1}} \dots \tilde {\cal D}_{\mu_{s-\frac{1}{2}+k+1}} \Psi_{\mu_1 \dots \mu_{s-\frac{1}{2}})^T}\,. \label{shiftinvfieldstrengthDde}
\ee
This is the basic local operator capturing all the on-shell non-trivial shift-invariant information of the theory.

\subsection{Mixed-symmetry massive fermions}

In higher dimensions, fermions can be in representations other than symmetric tensors.  The most general mixed-symmetry fermion \cite{Metsaev:1998xg,Zinoviev:2009vy,Skvortsov:2010nh,Reshetnyak:2012ec} is described by a tensor-spinor field in the representation $[s_1,s_2,\ldots,s_p]_{\rm s}$, i.e., the tensor indices live in the Young tableau with $p$ rows where the $i$-th row has length $s_i$.  (Note that $s$ in the previous sections referred to the spin of the fermion rather than the length of its tensor-spinor tableaux, so this was the case of a single row tableau with length $s_1=s-1/2$.)  The field is also divergenceless and gamma-traceless in every tensor index and satisfies the Dirac equation, 
\be
 \left(\slashed{\cal D}+\tilde m \right) \Psi_{\mu_1\ldots\mu_{s_1},\ldots} =0\,. 
 \ee
This implies that it also satisfies the Klein-Gordon equation,
\be \left({\cal D}^2 +{1\over L^2}\left[{D(D-1)\over 4}+\sum_i s_i\right]-\tilde m^2\right)  \Psi_{\mu_1\ldots\mu_{s_1},\ldots} =0\,,\ee
and is fully traceless in its tensor indices.

The AdS/CFT formula relating $\tilde m$ and the scaling dimension of the dual CFT operator is,
 \be \tilde m^2 L^2 =\left(\Delta-{d\over 2 }\right)^2\,,\ \ \Delta_\pm= {d\over 2}\pm \tilde m L\,,\label{mixedsymmasmrele}\ee
which is independent of the tableau of the tensor-spinor.

\subsection{Mixed-symmetry partially massless fermions}

The pattern of possible PM points for mixed-symmetry fermions mirrors that of their bosonic counterparts \cite{Metsaev:1995re,Metsaev:1997nj,Alkalaev:2003qv,Boulanger:2008up,Boulanger:2008kw,Skvortsov:2009zu,Skvortsov:2009nv,Basile:2016aen}.  
The partially massless points occur when squares in the field's tableau from a row that is longer than the row directly below it are `activated'.   If it is the $q$-th row that is activated, the number of squares that may be activated ranges over $1,2,\ldots, s_q-s_{q+1}$.  
We assign an integer depth label $\tilde t \in \{1,\cdots, s_q-s_{q+1} \}$, which indicates that $\tilde t$ of the squares in the $q$-th row are being activated. 
By removing these activated squares, we get the tableau corresponding to the tensor indices of the tensor-spinor gauge parameter, and each activated square becomes a $\tilde {\cal D}_\mu$ in the gauge transformation law.  
These PM points occur at half-integer values of the dual conformal dimension, given by
\be \Delta_+= d - q + s_{q}+{1\over 2} - \tilde t\,,\ee
which correspond to the mass values
\be \tilde m^2 L^2 =\left({D\over 2 }- q + s_{q} - \tilde t\right)^2 \,.\ee
They have a value of $\Delta_+$ that is larger by ${1/2}$ than that of their bosonic counterparts. 

There is a CFT unitarity bound $\Delta\geq d+s_1-h_1-{1\over 2}$ for a mixed-symmetry fermionic state \cite{Metsaev:1997nj,Brink:2000ag,Costa:2014rya}, where $h_1$ is the 
number of rows with length $s_1$.  This bound is saturated by the PM state where one square in the top block is activated, and all the other PM points lie below this bound.  Thus, on AdS space only this PM point is unitary and the rest are non-unitary.  On dS space, the condition for unitarity is more complicated and stems from the representation theory of the dS group: for $D$ odd, none of the fermionic PM points are unitary.  For $D$ even, the only unitary fermionic PM fields are those with a $(D-2)/2$ row tableau in which squares in the bottom row are activated \cite{10.1063/1.1665471}. 

If the tableaux has more than one row, $q>1$, the gauge transformation is reducible, and there are $q-1$ levels of gauge-for-gauge parameters.  As in the bosonic case discussed in Ref.~\cite{Hinterbichler:2022vcc}, the shift-symmetric fields will come as the longitudinal modes of massive fields as they approach partially massless limits with no reducibilities, i.e., those where the first row is activated.

Perhaps the most familiar example of fermions beyond the symmetric tensor-spinors are the fermionic $p$-forms \cite{Metsaev:1997hi,Buchbinder:2009pa,Zinoviev:2009wh,Campoleoni:2009gs,Lekeu:2021oti,Wang:2023iqt}, described by a tableau with one column of length $p$.  These have a single PM point where the bottom square is activated, which occurs at the value
\be \Delta_+= d - p +{1\over 2} \,,\ee
corresponding to the mass value
\be \tilde m^2L^2=\left({D\over 2}-p\right)^2\,.\ee
Since this is the only PM point for a fermionic $p$-form, we often call this the massless $p$-form and define the physical mass $m$ via
\be\tilde m^2={1\over L^2}\left( {D\over 2}-p\right)^2+ m^2\,,\label{mixedsymassrelepforme}\ee
so that $m=0$ corresponds to the massless point.

\subsection{Mixed-symmetry shift-symmetric fermions}

We find the shift-symmetric fermions by following the bosonic discussion of Ref.~\cite{Hinterbichler:2022vcc}: consider a PM fermion with tableaux $[s_1+k+1,s_2,\ldots,s_p]_{\rm s}$ where $k+1$ boxes in the first row are activated ($\tilde t=k+1$). 
These are the only PM fermions with irreducible gauge symmetries where the gauge parameter is a fermion of type $[s_1,s_2,\ldots,s_p]_{\rm s}$.  These PM points occur at the values
\be \Delta_+=d +  s_1-{1\over 2}\,.\ee
The gauge symmetry at this PM point has $k+1$ derivatives and so the dual operator has a conservation-type shortening condition at level $k+1$, yielding a null state fermion with tableaux $[s_1,s_2,\ldots, s_p]_{\rm s}$ and conformal dimension 
\be \Delta_+=d +k  +  s_1+{1\over 2}.\label{genmixwsconfde}\ee  
As this PM value is approached, the representation $D(\Delta,[s_1+k+1,s_2,\ldots, s_p]_{\rm s})$ splits via the branching rule
\begin{align}
D\!\left(\Delta,[s_1+k+1,s_2,\ldots, s_p]_{\rm s}\right) \xrightarrow[\Delta\rightarrow d  +  s_1-{1\over 2} ]{} & D\!\left( d +  s_1 -{1\over 2} ,[s_1+k+1,s_2,\ldots, s_p]_{\rm s}\right)  \nn \\
& \oplus D \! \left(d +k  +  s_1+{1\over 2},[s_1,s_2,\ldots, s_p]_{\rm s}\right)\,. 
\end{align}
The ``longitudinal'' representation $D(d +k  +  s_1+{1\over 2},[s_1,s_2,\ldots, s_p]_{\rm s})$ that splits off is the level-$k$ shift-symmetric fermion.  Using \eqref{mixedsymmasmrele} we find the masses for the shift-symmetric fermions of type $[s_1,\ldots, s_p]_{\rm s}$, 
\be \tilde{\mathsf{m}}^2_{[s_1,\ldots, s_p]_{\rm s},k}L^2=\left({D\over 2}+s_1+k\right)^2  \,. \ \label{gensymshsmassee}\ee
On AdS space, these shift-symmetric points all lie above the unitarity bound and hence are unitary.  On dS space, they are non-unitary.

The form of the shift symmetry is given by a fermionic ambient space tensor-spinor of type $[s_1+k,s_1,s_2,\ldots,s_p]_{\rm s}$, where the indices of the top row are contracted with ambient space coordinates $X^A$ and the rest are projected down.
 This can be thought of as the most general kind of spinorial Killing--Yano-like object on AdS$_D$.
 
 Each shift-symmetric field has a corresponding $(k+1)$-derivative fermionic `field strength' with the tableau $[s_1+k+1,s_2,\cdots,s_p]_{\rm s}$ of the parent PM field,
 \be F_{\mu_1\cdots\mu_{s_1+k+1},\cdots} = {\cal Y}_{{[s_1+k+1,s_2,\cdots,s_p]}^T} \left[ \tilde {\cal D}_{\mu_{s_1+1}}\cdots \tilde {\cal D}_{\mu_{s_1+k+1}} \Phi_{\mu_1\cdots\mu_{s_1},\cdots}  \right] \label{genmixsyfise},\ee
 where $ {\cal Y}$ is a Young projector.
 This field strength is invariant under the shift symmetries and is the basic local operator capturing all of the on-shell non-trivial shift-invariant information of the theory.

\section{Ambient space and bispinors}
\label{sec:ambient}

In this section, we will specialize to the case of AdS$_4$. 
We discuss the notation and kinematics necessary to describe fermionic fields on AdS$_4$ using ambient/embedding space.  We make use of the ambient space approach of Ref.~\cite{Pethybridge:2021rwf}  and the bispinor formalism of Ref.~\cite{Binder:2020raz}, which exploits the isomorphism between the isometry algebra $\mathfrak{so}(2,3)$ of AdS$_4$ and the symplectic algebra $\mathfrak{sp}(4, \mathbb{R})$.  This formalism is analogous to the spinor-helicity formalism often used for flat space, and it will allow us to treat all spins, even and odd, in a uniform manner.

\subsection{Spinor conventions}
\label{subsec:spinors}

There are many index types to keep track of when dealing with fermions in ambient space, so we first list these and specify our conventions.
 
 \begin{itemize}
 \item
The indices $\mu,\nu,\ldots$ are intrinsic spacetime indices for AdS$_4$ with coordinates $x^{\mu}$.  Spacetime indices are raised and lowered with the metric $g_{\mu \nu}$.
\item
The indices $i,j,\ldots$ are for the 3D Minkowski space $\mathbb{R}^{1,2}$.  These indices are raised and lowered with the 3D Minkowski metric $\eta_{ij}$.  
\item
The indices $a,b,\ldots$ are local Lorentz tangent space indices for AdS$_4$.  The 4D tangent space vector indices are raised and lowered with the  local Minkowski metric $\eta_{ab}$.   The vierbein is $e_\mu^{ a}$.  
\item
The indices $\alpha, \beta, \dots$, $\dot{\alpha}, \dot{\beta}, \dots$ are $SL(2, \mathbb{C})$ spinor indices for the tangent space.   These indices are raised and lowered using the 2D  invariant anti-symmetric tensors $\epsilon_{\alpha \beta}$, $\epsilon_{\dot{\alpha} \dot{\beta}}$ and their raised versions $\epsilon^{\alpha \beta}$, $\epsilon^{\dot{\alpha} \dot{\beta}}$. Our conventions for raising and lowering indices are given by
\be
s_{\alpha} = \epsilon_{\alpha \beta} s^{\beta}, \quad s^{\alpha} = s_{\beta} \epsilon^{\beta \alpha}. 
\ee
With this convention, $\epsilon^{\alpha \beta}$ is equal to  $\epsilon_{\alpha \beta}$ with raised indices and the Kronecker delta is $\delta^{\alpha}{}_{\beta} =\epsilon^{\alpha \gamma} \epsilon_{\beta \gamma}=-\delta_{\beta}{}^{\alpha}$, and similarly for dotted indices.
The 4D sigma matrices are $\sigma^{a \alpha \dot{\alpha}}$ and $\bar{\sigma}^a_{\dot{\alpha} \alpha}$. From the vierbein $e_{\mu}^a$ and its inverse $e^{\mu}_a$, we define $\sigma_{\mu}^{ \alpha \dot{\alpha}}\coloneqq e_{\mu}^a \sigma^{a \alpha \dot{\alpha}}$ and $\bar{\sigma}^{\mu}_{\dot{\alpha} \alpha}\coloneqq e^{\mu}_a\bar{\sigma}^a_{\dot{\alpha} \alpha}$.
\item
The indices $A,B,\ldots$ are for the Cartesian coordinates $X^A$ of the ambient space $\mathbb{R}^{3,2}$ which has the flat metric $\eta_{AB}$, used to raise and lower indices on ambient space tensors.
\item
The indices $I,J,\ldots$ are $Sp(4, \mathbb{R})$ spinor indices for the ambient space.  The invariant anti-symmetric tensor is $J_{IJ}$.  The inverse transpose of the antisymmetric tensor is denoted by $J^{IJ}$.
We raise and lower indices using $J_{IJ}$ and $J^{IJ}$ as
\be
S_I = J_{IJ} S^J, \quad S^I = S_J J^{JI}.
\ee
With this convention, $J^{IJ}$ is equal to $J_{IJ}$ with raised indices, i.e., $J^{IJ} = J_{KL}J^{KI}J^{LJ}$, and the Kronecker delta is $\delta^{I}{}_J =J^{IK}J_{JK} = - \delta_J{}^I$.  The relative position of contracted indices is important, e.g., we have $S_{I}{}^{I \dots} = - S^I{}_{I}{}^{\dots}$.
\end{itemize}

The different indices and invariant tensors are summarized in Table~\ref{table:conventions}.
\begin{table}[ht]
\begin{center}
\begin{tabular}{ |c | c | c|c| }
\hline
 Type & Indices & Range & Invariant tensors  \\ 
 \hline 
 AdS$_4$ spacetime & $\mu, \nu, \dots$ & $0,\dots, 3$ & $g_{\mu \nu}$\\ 
  $SO(1,2)$ & $i, j, \dots$ & $0,\dots,2$  & $\eta_{ij}$ \\  
  $SO(1,3)$ & $a, b, \dots$ & $0,\dots, 3$  & $\eta_{ab}$\\  
   $SL(2, \mathbb{C})$ &  $\alpha, \beta, \dots$, $\dot{\alpha}, \dot{\beta}, \dots$  &$1, 2$ & $\epsilon_{\alpha \beta}, \epsilon_{\dot{\alpha}, \dot{\beta}}$ \\
  $SO(2,3)$ & $A, B, \dots$ & $0, \dots, 4$ & $\eta_{AB}$ \\
  $Sp(4, \mathbb{R})$ & $I, J, \dots$ & $1, \dots, 4$& $J_{IJ}$ \\
  \hline
\end{tabular}
\caption{Summary of different indices and invariant tensors.}
\label{table:conventions}
\end{center}
\end{table}

The five $SO(2,3)$ gamma matrices are denoted by $(\Gamma^A)^I{}_J$. They satisfy the Clifford algebra anticommutation relations,
\be
\{ \Gamma^A, \Gamma^B \} = 2 \eta^{AB}.
\ee
They can be chosen to be real and unitary.
Taking $\eta^{AB} = {\rm diag}(-1, 1, 1, 1, -1)$, a convenient such basis is the one used by Ref.~\cite{Binder:2020raz} with
\be
J_{IJ} = 
\begin{pmatrix}
 0 & 0 & 1 & 0 \\
 0 & 0 & 0 & 1 \\
 -1 & 0 & 0 & 0 \\
 0 & -1 & 0 & 0 \\
\end{pmatrix}
\ee
and
\begin{align}
& (\Gamma^0)^I{}_J = \begin{pmatrix}-i \sigma^2 & 0 \\ 0 & i \sigma^2\end{pmatrix},  \quad
(\Gamma^1)^I{}_J = \begin{pmatrix} \sigma^1 & 0 \\ 0 & \sigma^1\end{pmatrix}, \quad 
(\Gamma^2 )^I{}_J = \begin{pmatrix} \sigma^3 & 0 \\ 0 & \sigma^3\end{pmatrix}, \\
& (\Gamma^3)^I{}_J = \begin{pmatrix}0 & i  \sigma^2 \\ -i  \sigma^2& 0 \end{pmatrix}, \quad (\Gamma^4)^I{}_J = \begin{pmatrix}0 &- i  \sigma^2 \\ -i  \sigma^2& 0 \end{pmatrix},
\end{align}
where
\be \label{eq:sigma-matrices}
\sigma^1 = \begin{pmatrix} 0 & 1 \\ 1 &0 \end{pmatrix}, \qquad \sigma^2 = \begin{pmatrix} 0 & -i \\ i  & 0 \end{pmatrix}, \qquad  \sigma^3 = \begin{pmatrix} 1 & 0 \\ 0  & -1 \end{pmatrix}.
\ee
The corresponding 4D sigma matrices $\sigma_a^{\alpha \dot{\alpha}}$ are
\be \label{eq:4Dsigma}
\sigma_0^{\alpha \dot{\alpha}} =\begin{pmatrix} -1 & 0 \\ 0 & -1 \end{pmatrix}, \qquad \sigma_1^{\alpha \dot{\alpha}} =\begin{pmatrix} -1 & 0 \\ 0 & 1 \end{pmatrix}, \qquad \sigma_2^{\alpha \dot{\alpha}} =\begin{pmatrix} 0 & 1 \\ 1 & 0 \end{pmatrix}, \qquad \sigma_3^{\alpha \dot{\alpha}} =\begin{pmatrix} 0 & i \\ -i & 0 \end{pmatrix},
\ee
the 4D gamma matrices $\gamma^a$ are
\be
\gamma^a =\begin{pmatrix} 0 & i \sigma^{a}_{\alpha \dot{\alpha}} \\ i (\bar{\sigma}^a)^{ \dot{\alpha} \alpha}  & 0 \end{pmatrix},
\ee
where $(\bar{\sigma}^{a})^{ \dot{\alpha} \alpha} =(\sigma^{a T})^{\alpha \dot{\alpha}}$, and the 3D gamma matrices $\gamma_{(3)}^i$ are 
\be \label{eq:3D-gamma} 
\gamma_{(3)}^0 = -i \sigma^2, \qquad \gamma_{(3)}^1 = \sigma^1, \qquad \gamma_{(3)}^2 = \sigma^3.
\ee
For a vector $P_A$ we define $P^I{}_J \coloneqq P_A(\Gamma^A)^I{}_J$, where $P_{IJ}$ is antisymmetric and $J$-traceless. The inverse relation is $P^A = \frac{1}{4} P^I{}_J (\Gamma^A)^J{}_I$. 
For spinors $S^I$, $T^I$ and  $P_1^I{}_J , \dots, P_n^I{}_J $, we use the angle bracket notation
\be
\langle S P_1 \dots P_n T\rangle \coloneqq S_{I_1}P^{I_1}{}_{I_2} \dots  P^{I_{2n-1}}{}_{I_{2n}} T^{I_{2n}}.
\ee
For the bispinors $T_I^{\alpha}$, $\bar{T}_I^{\dot{\alpha}}$, we also use angle brackets to denote contractions of the $SL(2, \mathbb{C})$ indices, 
\be
 \langle T^I T^J \rangle \coloneqq T^I_{\alpha} T^{\alpha J}, \quad  \langle \bar{T}^I \bar{T}^J \rangle \coloneqq \bar{T}^I_{\alpha} \bar{T}^{\alpha J}, \quad \langle T T \rangle \coloneqq T_{\alpha I} T^{\alpha I}, \quad \langle \bar{T} \bar{T} \rangle \coloneqq \bar{T}_{\dot{\alpha} I} \bar{T}^{\dot{\alpha} I}.
\ee

\subsection{Representations}
The group $SL(2, \mathbb{C})$ is the double cover of $SO_0(1,3)$, the connected component of the identity of $SO(1,3)$, while $Sp(4, \mathbb{R})$ is the double cover of $SO_0(2,3)$.
It will be useful to review the Lie algebra $\mathfrak{so}(2,3) \cong \mathfrak{sp}(4, \mathbb{R})$ and the transformations from tensor representations of the indefinite orthogonal groups to representations of their covers. 

\subsubsection{$\mathfrak{so}(2,3) \cong \mathfrak{sp}(4, \mathbb{R})$}

The 10-dimensional Lie algebra $\mathfrak{sp}(4, \mathbb{R})$ consists of all real $4\times 4$ matrices $M^I{}_J$ preserving the symplectic form $J_{IJ}$,
\be
M^K{}_I J_{K J} + M^K{}_J J_{I K} = 0.
\ee
We can write a basis of such matrices as $(M^{AB})^I{}_J =\frac{1}{2}(\Gamma^{AB})^I{}_J$,  where $\Gamma^{AB} \coloneqq \frac{1}{2}[\Gamma^A, \Gamma^B]$. These satisfy the commutation relations of $\mathfrak{so}(2,3)$,
\be \label{eq:conformal-algebra}
[M_{AB}, M_{CD}]=-\eta_{AC} M_{BD}+\eta_{BC} M_{AD}-\eta_{BD} M_{AC}+\eta_{AD} M_{BC}.
\ee
For later use, we define the conformal algebra generators $D \coloneqq M_{45}$, $P_i \coloneqq M_{i 4}-M_{i5}$, $K_i \coloneqq M_{i 4}+M_{i5}$, and $m_{ij} \coloneqq M_{ij}$, with $i, j=1,2,3$,  so the non-vanishing commutators in \eqref{eq:conformal-algebra} can be written as
\begin{subequations} \label{eq:conformal-algebra-2}
\begin{align}
[m_{ij}, m_{kl} ]  & = -\eta_{ik} m_{jl}+\eta_{jk} m_{il} -\eta_{jl} m_{ik} + \eta_{il} m_{jk}, \\
[m_{ij}, P_k] &= -\eta_{ik} P_j+\eta_{jk} P_i, \\
[m_{ij}, K_k] &= -\eta_{ik} K_j+\eta_{jk} K_i, \\
[K_i, P_j ] & = 2 \eta_{ij} D-2 m_{ij} , \\
[D, P_i ] & = P_i,  \\
[D, K_i ]  & = -K_i.
\end{align}
\end{subequations}

\subsubsection{$SL(2, \mathbb{C})$ representations}
Let us briefly review the relation between representations of $SO(1,3)$ and $SL(2, \mathbb{C})$.
We have the completeness relations involving the 4D sigma matrices \eqref{eq:4Dsigma}, 
\be
\sigma_a^{\alpha \dot{\alpha}} \bar{\sigma}^a_{\dot{\beta} \beta} = { -}2 \delta^{\alpha}{}_{\beta} \delta^{\dot{\alpha}}{}_{\dot{\beta}},\quad  (\sigma^{a})^{\alpha \dot{\alpha}} \sigma_a^{ \beta \dot{\beta}} = -2 \epsilon^{\alpha \beta} \epsilon^{\dot{\alpha} \dot{\beta}}, \quad \bar{\sigma}^a_{ \dot{\alpha} \alpha} \bar{\sigma}_{a \dot{\beta} \beta } = -2 \epsilon_{\alpha \beta} \epsilon_{\dot{\alpha} \dot{\beta}},
\ee
\be
(\sigma^a)^{\alpha \dot{\alpha}} \sigma^b_{\alpha \dot{\alpha}} = -2 \eta^{ab}, \quad (\sigma^a)^{\alpha \dot{\alpha}} \bar{\sigma}^b_{\dot{\alpha} \alpha } = -2 \eta^{ab}, \quad (\bar{\sigma}^a)^{ \dot{\alpha} \alpha} \bar{\sigma}^b_{\dot{\alpha} \alpha } = -2 \eta^{ab}.
\ee
We can map a vector $v^{a}$ of $SO(1,3)$ to the $SL(2, \mathbb{C})$ representation $v^{\alpha \dot{\alpha}}$ using
\be \label{eq:vector-map}
v^{\alpha \dot{\alpha}}=  \sigma_{a}^{ \alpha \dot{\alpha}} v^{a}, \quad v^{a}= { -}\frac{1}{2} { {\sigma}^{a}_{\alpha \dot{\alpha}}} v^{\alpha \dot{\alpha}}.
\ee
This implies that $v_a v^a = -\frac{1}{2} v^{\alpha \dot{\alpha}} v_{\alpha \dot{\alpha}}$.
More generally, we can map a symmetric traceless rank-$s$ tensor $v^{a_1 \dots a_s}$ of $SO(1,3)$ to the $SL(2, \mathbb{C})$ representation $v^{\alpha_1 \dots \alpha_s, \dot{\alpha}_1 \dots \dot{\alpha}_s}$ that is totally symmetric in the undotted and dotted indices,
\be
v^{\alpha_1 \dots \alpha_s, \dot{\alpha}_1 \dots \dot{\alpha}_s} = \sigma_{a_1}^{\alpha_1 \dot{\alpha}_s}\dots \sigma_{a_s}^{\alpha_1 \dot{\alpha}_s} v^{a_1 \dots a_s}, \quad v^{a_1 \dots a_s} = \frac{{(-1)^s} }{2^s} { {\sigma}^{a_1}_{\alpha_1 \dot{\alpha}_1}\dots {\sigma}^{a_s}_{\alpha_s \dot{\alpha}_s}} v^{\alpha_1 \dots \alpha_s, \dot{\alpha}_1 \dots \dot{\alpha}_s}.
\ee
We also have the more general finite-dimensional representations of $SL(2, \mathbb{C})$,
\be
v^{\alpha_1 \dots \alpha_{m}, \dot{\alpha}_1 \dots \dot{\alpha}_{n}},
\ee
where the undotted and dotted indices are each completely symmetric.
\subsubsection{$Sp(4, \mathbb{R})$ representations}

We now discuss the relation between finite-dimensional representations of $SO(2,3)$ and $Sp(4,{\mathbb R})$.
Raising an index on the gamma matrices, the result is anti-symmetric and $J$-traceless,
\be 
\left(\Gamma^A\right)^{IJ}=\left(\Gamma^A\right)^{I}_{\ K}J^{KJ},\quad \left(\Gamma^A\right)^{IJ}=-\left(\Gamma^A\right)^{JI}\, , \quad \left(\Gamma^A\right)^{IJ}J_{IJ}=0 \,.
\ee
We have the completeness relations
\be 
\label{eq:completeness1}
 \left(\Gamma^A\right)^{IJ}\left(\Gamma^B\right)_{IJ}=4\eta^{AB}, \quad \left(\Gamma^A\right)^{IJ}\left(\Gamma_A\right)^{KL}=2\left(J^{IK}J^{JL}-J^{IL}J^{JK}\right)-J^{IJ}J^{KL}. 
\ee
Using this, we can map an $SO(2,3)$ vector $V^A$ to an anti-symmetric $J$-traceless $Sp(4,{\mathbb R})$ tensor $V^{IJ}$,
\be 
V^{IJ}=V_A \left(\Gamma^A\right)^{IJ}\,, \quad V^A={1\over 4}  \left(\Gamma^A\right)_{IJ} V^{IJ} .
\ee
We have the convenient identity
\be
V^I{}_J V^J{}_K = V^2 \delta^I{}_K, 
\ee
where $V^2 \coloneqq \eta_{AB} V^A V^B = \frac{1}{4} V_{IJ} V^{IJ}$.

Raising an index on the generators $\Gamma_{AB}$, the result is symmetric,
\be 
\left(\Gamma_{AB}\right)^{IJ}=\left(\Gamma_{AB}\right)^{I}{}_{K}J^{KJ}, \quad \left(\Gamma_{AB}\right)^{IJ}=\left(\Gamma_{AB}\right)^{JI}  \, .
\ee
We have the completeness relations
\be 
\left(\Gamma^{AB}\right)^{IJ}\left(\Gamma^{CD}\right)_{IJ}=4\left(\eta^{A C}\eta^{BD}-\eta^{AD}\eta^{BC}\right), \quad \left(\Gamma^{AB}\right)^{IJ}\left(\Gamma_{AB}\right)^{KL}= 4\left(J^{IK}J^{JL}+J^{IL}J^{JK}\right).
\label{gammagammacome}
\ee
Using this, we can map a rank-2 anti-symmetric $SO(2,3)$ tensor $T^{AB}$ to a rank-2 symmetric $Sp(4,{\mathbb R})$ tensor $T^{IJ}$,
\be 
T^{IJ}={1\over 2}T_{AB} \left(\Gamma^{AB}\right)^{IJ}\,, \quad T^{AB}={1\over 4}  \left(\Gamma^{AB}\right)_{IJ} T^{IJ} .\label{ansytosyme}
\ee

In general, an $SO(2,3)$ irreducible tensor representation is specified by a two-row traceless Young tableau $[m_1,m_2]$.  We can include the spinor representations by including a formal fundamental spinor tableau $[\frac{1}{2},\frac{1}{2}]$, so that the irreducible tensor-spinor representation with tensor indices $[m_1,m_2]$ and a spinor index is denoted by $[m_1+\frac{1}{2},m_2+\frac{1}{2}]$ (note that this differs from the notation $[\cdot, \cdot ]_{\rm s}$ used in the previous section).  Including these representations accounts for all irreducible representations. 
The representations are all real. The $SO(2,3)$ representation $[m_1,m_2]$ with $2m_i \in \mathbb{Z}_{\geq 0}$ is equivalent to the $Sp(4,{\mathbb R})$ representation given by the two-row $J$-traceless tableau $[m_1+m_2,m_1-m_2]$, 
\be
\raisebox{1.15ex}{\gyoung(_7{m_1+m_2},_5{m_1 -m_2})} \, .
\ee 
Those $Sp(4,{\mathbb R})$ representations given by tableaux with an odd number of boxes are fermionic representations, while those with an even number of boxes are bosonic representations. For example, the spin-$s$ $SO(2,3)$  representation, carried by a symmetric traceless rank-$s$ tensor, is described by the $Sp(4,{\mathbb R})$ tensor with the symmetries of the $J$-traceless tableau $[s, s]$.

\subsection{Ambient space}

Now we discuss the ambient space construction for fermionic fields on AdS$_4$. The original bosonic ambient space goes back to Ref.~\cite{Dirac-ambient}. For a review, see Ref.~\cite{Bekaert:2010hk}. To describe fermions in ambient space, we adapt to AdS$_4$ the even-dimensional dS formalism from Ref.~\cite{Pethybridge:2021rwf} using the bispinor formalism of Ref.~\cite{Binder:2020raz}. See also Refs.~\cite{Takook:2014paa, Iliesiu:2015qra, Nishida:2018opl, Schaub:2023scu}. 

\subsubsection{Ambient space}
Consider the ambient space $\mathbb{R}^{2,3}$ with coordinates $X^A$ with $A=0, \dots, 4$ and metric $\eta_{AB} = {\rm diag}(-1, 1, 1, 1, -1)$. Define the open submanifold 
\be
\mathbb{R}_0^{2,3} = \{ X^A:  X^2 \coloneqq \eta_{AB} X^A X^B <0 \} \subset \mathbb{R}^{2,3}.
\ee 
We identify AdS$_4$ with the universal cover of the hypersurface in $\mathbb{R}_0^{2,3}$ defined by $X^2= - L^2$. If $x^{\mu}$ are local coordinates on AdS$_4$, then we have an embedding $x^{\mu} \mapsto X^A(x)$, so we have
\be  \label{eq:AdS-constraint}
X_A(x) X^A(x)= -L^{2}.
\ee 
For example, in Poincar\'e coordinates $x^{\mu} = (x^i, z)$ we can take
\be
X^i (x)= \frac{L x^i}{z}, \quad i=0,1,2, \quad  X^3(x) = \frac{L^2-\eta_{ij} x^i x^j -z^2}{2z}, \quad X^4(x) = \frac{L^2+\eta_{ij} x^i x^j +z^2}{2z}, 
\ee
where $\eta_{ij} = {\rm diag}(-1, 1, 1)$, and the induced metric on AdS$_4$ is
\be
ds^2 = \frac{L^2}{z^2}(\eta_{ij} d x^i dx^j +dz^2).
\ee

From the ambient space coordinates $X^A$, we can define the anti-symmetric $J$-traceless $Sp(4, \mathbb{R})$ tensor $X^{IJ}$,
\be 
X^{IJ}=X_A \left(\Gamma^A\right)^{IJ}\,, \quad X^A={1\over 4}  \left(\Gamma^A\right)_{IJ} X^{IJ}.
\ee
It is also useful to define the following projectors in ambient space:
\be \label{eq:projectors}
\mathcal{P}_{\pm}^I{}_J(X) \coloneqq \frac{1}{2} \left(\delta^I{}_J \mp \frac{i}{\sqrt{-X^2}} X^I{}_J \right).
\ee
These project onto the spaces of eigenspinors of the coordinate matrix $X^I{}_J$,
\be
X^I{}_J \mathcal{P}_{\pm}^J{}_K = \pm i \sqrt{-X^2}\mathcal{P}_{\pm}^I{}_K,
\ee
and satisfy the conditions
\be
\mathcal{P}_{+}^I{}_J +\mathcal{P}_{-}^I{}_J=\delta^I{}_J, \qquad \mathcal{P}_{\pm}^I{}_J \mathcal{P}_{\pm}^J{}_K  = \mathcal{P}_{\pm}^I{}_K, \qquad \mathcal{P}_{+}^I{}_J \mathcal{P}_{-}^J{}_K = 0,
\ee
where here the dependence of the projectors on the coordinates $X^A$ is left implicit.
\subsubsection{Bispinors}

Following Ref.~\cite{Binder:2020raz}, we introduce the complex bispinors ${{T}}^I_{\alpha}$  on AdS$_4$, $I=1, \dots 4$ and $\alpha=1,2$, such that we can write the embedding coordinates $X^A(x)$ as
\be \label{eq:bispinor-def}
X^A(x) = - \frac{1}{4} \langle {{T}}_{\alpha} \Gamma^A {{T}}^{\alpha} \rangle \, .
\ee
The conjugate bispinor is $\bar{{T}}^I_{\dot{\alpha}} \coloneqq ({{T}}^I_{\alpha})^*$. The bispinor transforms covariantly under both $Sp(4, \mathbb{R})$ and $SL(2, \mathbb{C})$. Equation~\eqref{eq:bispinor-def} determines $T_{\alpha}^I$ in terms of $X^A(x)$ up to $SL(2, \mathbb{C})$ transformations. In terms of $X^{IJ}$, Equation~\eqref{eq:bispinor-def} is 
 \be \label{eq:Xslash}
X^{IJ}(x) = { T}^I_{\alpha} { T}^{\alpha J} - i L J^{IJ}, 
\ee
where we have used the second identity in \eqref{eq:completeness1}. Note that ${{T}}^I_{\alpha}$ is defined only on the AdS surface, rather than the full ambient space.

By Eqs.~\eqref{eq:AdS-constraint} and \eqref{eq:bispinor-def}, and the reality of $X^A(x)$, the bispinors satisfy the constraints
\be \label{eq:spinor-constraints}
\langle T_{\alpha} \Gamma^A T^{\alpha} \rangle =  \langle \bar{T}_{\dot{\alpha}} \Gamma^A \bar{T}^{\dot{\alpha}} \rangle, \quad \langle T T  \rangle^2  =\langle \bar T \bar T  \rangle^2= -16 L^{2}.
\ee
These represent six independent real constraints, since the first equality already implies that $ \langle T T  \rangle^2$ is real. Together with $SL(2, \mathbb{C})$ covariance, these six constraints reduce the 16 real degrees of freedom in $T^I_{\alpha}$ to four real degrees of freedom, corresponding to the four coordinates $x^{\mu}$.
The constraints can equivalently be written as \cite{Binder:2020raz}
\be \label{eq:AdS-on-shell}
\langle T^{\alpha} \bar{T}^{\dot{\alpha} } \rangle =0, \quad \langle T^{\alpha} T^{\beta } \rangle= 2i L\epsilon^{\alpha \beta}, \quad \langle \bar{T}^{\dot{\alpha}} \bar{T}^{\dot{\beta} } \rangle = -2i L\epsilon^{\dot{\alpha} \dot{\beta}}.
\ee
This equivalence is not obvious, but it can be checked by comparing Gr\"obner bases of each system of equations using an explicit choice of $\Gamma^A$ and $J_{IJ}$. 

From \eqref{eq:Xslash}, we see that the bispinors are eigenspinors of the coordinate matrix, 
\be \label{eq:eigenspinor}
X^{I}{}_J(x) T^J_{\alpha} = i L T^I_{\alpha}.
\ee
In terms of the projectors \eqref{eq:projectors}, this reads
\be \label{eq:eigenspinor-2}
\mathcal{P}_{-}^I{}_J(X(x)) T^J_{\alpha} =0.
\ee
For example, using Poincar\'e coordinates and the explicit $\Gamma^A$ from Section~\ref{subsec:spinors}, we can write
\be
T^I_{\alpha} = \frac{1}{\sqrt{z}}\begin{pmatrix} L & 0 \\ 0 & L \\ -x^0+x^1 & -i z -x^2 \\ iz -x^2 & -x^0-x^1 \end{pmatrix}.
\ee

The bispinors $T^I_{\alpha}$ give the full set of Killing spinors on AdS$_4$ \cite{Binder:2020raz}. If we form the Dirac spinors $\xi^I = \begin{pmatrix} T_{\alpha}^I \\ \bar{T}^{\dot{\alpha} I} \end{pmatrix}$, then these satisfy the AdS$_4$ Killing spinor equations, 
\be
\left({\cal D}^{\mu} - \frac{1}{2L} \gamma^{\mu} \right) \xi^I =0.
\ee
In terms of Weyl spinors and ${\cal D}_{\alpha \dot{\alpha}} = \sigma^{\mu}_{\alpha \dot{\alpha}} {\cal D}_{\mu}$, these Killing spinor equations are
\be
{\cal D}_{\alpha \dot{\alpha}} T_{\beta}^I =  -\frac{i}{L} \epsilon_{ \alpha \beta}  \bar{T}_{\dot{\alpha}}^I, \quad {\cal D}_{\alpha \dot{\alpha}} \bar{T}_{\dot{\beta}}^I= \frac{i}{L} \epsilon_{\dot{\alpha} \dot{\beta}}T_{ \alpha}^I .
\ee
From these we can deduce that the covariant derivative acting on Killing spinors can be written as
\be \label{eq:cd}
{\cal D}^{\alpha \dot{\alpha}} =  \frac{i}{L} \left( T^{\alpha I} \bar{\partial}_I^{\dot{\alpha}} - \bar{T}^{\dot{\alpha} I} \partial_I^{\alpha}\right),
\ee
where we use the notation $\partial_I^{ \alpha} \coloneqq \partial/\partial T^I_{\alpha}$ and $\bar{\partial}_I^{\dot{\alpha}} \coloneqq \partial/\partial \bar{T}^I_{\dot{\alpha}}$.

\subsubsection{Vierbein}
We can write the AdS$_4$ metric $g_{\mu \nu}$ in terms of the bispinors by projecting the flat ambient metric,
\be
g_{\mu \nu} = \eta_{AB} \partial_{\mu} X^A(x) \partial_{\nu} X^B(x) = -\frac{1}{2} \langle T^{\alpha} \partial_{\mu} \bar{T}^{\dot{\alpha}} \rangle \langle T_{\alpha} \partial_{\nu} \bar{T}_{\dot{\alpha}} \rangle,
\ee
where $\partial_{\mu} \coloneqq \partial/\partial x^{\mu}$ and we have used the completeness relation \eqref{eq:completeness1} and the on-shell conditions \eqref{eq:AdS-on-shell}. 
Following Ref.~\cite{Binder:2020raz}, the vierbein can be written as
\be
e^{a}_{\mu} = \frac{1}{2} (\sigma^{a})^{\alpha \dot{\alpha} } \langle T_{\alpha} \partial_{\mu} \bar{T}_{\dot{\alpha}} \rangle ,
\ee
and the projector onto the AdS surface can be written as 
\be \label{eq:projector}
\partial_{\mu} X^A = \frac{i}{4 L} e_{\mu}^a \sigma_a^{\alpha \dot{\alpha}}  \langle T_{\alpha} \Gamma^A \bar{T}_{\dot{\alpha}} \rangle.
\ee

\subsubsection{Projecting tensors and spinors}
We now discuss the ambient space representations of different types of AdS$_4$ fields. We first consider scalars and spinors, then discuss general symmetric tensors and tensor-spinors.

\subsubsection*{Scalars}
Given a real ambient space scalar field $\Phi(X)$, we project it to an AdS scalar field $\phi(x)$ by simply restricting to the AdS surface, 
\be
\Phi( X) \mapsto \phi(x)= \Phi( X(x)). 
\ee
To make this map a bijection we restrict $\Phi(X)$ to have definite homogeneity degree under real rescalings, i.e., for $\lambda \in \mathbb{R}$ we require 
\be
\Phi(\lambda X) = \lambda^{w} \Phi(X),
\ee
for some real constant $w$ called the weight. This then uniquely determines the ambient field away from the AdS surface in terms of its restriction $\phi(x)$.

\subsubsection*{Spinors}

Given an ambient Dirac spinor $\Phi_I(X)$, we can project it to a left-handed AdS$_4$ Weyl spinor $\phi_{\alpha}(x)$ by restricting to the AdS$_4$ surface and contracting with a bispinor,
\be \label{eq:spinor-projection}
\Phi_I(X) \mapsto \phi_{\alpha}(x) = \frac{1}{\sqrt{2L}} T_{\alpha}^I \Phi_I(X(x)),
\ee
where we have fixed the overall normalization (up to an arbitrary phase) by matching to the pullback of a vector \eqref{eq:pullback}.
To make this map a bijection we restrict $\Phi^I(X)$ to have definite homogeneity degree and to be an eigenspinor of the coordinate matrix $X^I{}_J$,
\be \label{eigenspinordconde}
\Phi^I(\lambda X) = \lambda^w \Phi^I(X), \quad \mathcal{P}_{-}^I{}_J(X) \Phi^J(X) =0.
\ee
This eigenspinor constraint is the AdS version of the equation considered in Ref.~\cite{Pethybridge:2021rwf}: it is necessary to trivialise the kernel of the projection, which would otherwise be spanned by terms that when restricted to AdS take the form $\bar{T}^I_{\dot{\alpha}} f^{\dot{\alpha}}(x)$, for some arbitrary spinor $f^{\dot{\alpha}}(x)$. We can ensure that any ambient spinor index satisfies the eigenspinor constraint equation in \eqref{eigenspinordconde} by contracting with $\mathcal{P}_{+}^I{}_J$. So given an AdS$_4$ spinor $\phi_{\alpha}(x)$, we find its ambient representative $\Phi_I(X)$  by writing
\be \label{eq:spinor-restriction}
\Phi^I(X(x))= \frac{i}{\sqrt{2L}} T^I_{\alpha} \phi^{\alpha}(x)
\ee
and using homogeneity to define $\Phi_I(X)$ away from the surface.  Using \eqref{eq:eigenspinor-2}, we can see that this $\Phi_I(X)$ satisfies the eigenspinor constraint equation. 

Similarly, there is a bijection between right-handed Weyl spinors $\bar{\phi}^{\dot{\alpha}}(x)$ and ambient Dirac spinors $\Phi^I(X)$ satisfying
\be
\Phi^I(\lambda X) = \lambda^w \Phi^I(X), \quad \mathcal{P}_{+}^I{}_J(X) \Phi^J(X) =0,
\ee
where the projection is
\be \label{eq:conjugate-spinor-project}
\Phi_I(X) \mapsto \bar{\phi}_{\dot{\alpha}}(x) = \frac{1}{\sqrt{2L}} \bar{T}_{\dot{\alpha}}^I \Phi_I(X(x)).
\ee
We can ensure that any ambient spinor index satisfies this eigenspinor constraint by contracting with $\mathcal{P}_{-}^I{}_J$.

To see that \eqref{eq:spinor-projection} is correct, we can check that $\phi^{\alpha}(x)$ transforms under the isometries as expected for a Weyl spinor \cite{Pethybridge:2021rwf}. Under the isometries of AdS$_4$, $\Phi^{I}(X)$ transforms infinitesimally as
\be
\delta_{M_{AB}}  \Phi^{I} = (X_A \partial_B -X_B \partial_A)  \Phi^{I} +  \frac{1}{2}(\Gamma_{AB})^I{}_J  \Phi^{J},
\ee
with the commutation relations $[\delta_{M_{AB}}, \delta_{M_{CD}}]\Phi^{I} =\eta_{AC} \delta_{M_{BD}} \Phi^{I} + \dots$.
Using Poincar\'e coordinates and the 5D gamma matrices from Section~\ref{subsec:spinors}, this implies the following transformations for $\phi^{\alpha}(x)$ under the generators from \eqref{eq:conformal-algebra-2}:
\begin{align}
\delta_D \phi^{\alpha} & =-x^{\mu} \partial_{\mu} \phi^{\alpha}, \\
\delta_{m_{ij}} \phi^{\alpha} & = (x_i \partial_j -x_j \partial_i) \phi^{\alpha} + \frac{1}{2} \epsilon_{ijk} (\gamma_{(3)}^k)^{\alpha}{}_{\beta} \phi^{\beta}, \\
\delta_{P_i} \phi^{\alpha} & = -\partial_i \phi^{\alpha} \\
\delta_{K_i} \phi^{\alpha} & = -L^{-1} \left[ 2 x_i x^{\mu} \partial_{\mu} \phi^{\alpha}  -\eta_{\mu \nu} x^{\mu} x^{\nu} \partial_i \phi^{\alpha} +\epsilon_{ijk} x^j (\gamma_{(3)}^k)^{\alpha}{}_{\beta} \phi^{\beta} - i z \eta_{ij} (\gamma_{(3)}^j)^{\alpha}{}_{\beta} \phi^{\beta} \right],
\end{align}
where $\gamma_{(3)}^i$ are the real 3D gamma matrices given in \eqref{eq:3D-gamma}. 

\subsubsection*{Tensors}
Given an ambient $[s,s]$-tensor $\Phi_{I_1 J_1, \dots, I_s J_s}(X)$,
\be \label{eq:[s,s]}
\Phi_{I_1 J_1, \dots, I_s J_s} \sim \raisebox{1.15ex}{\gyoung(_4{s},_4s)} \, ,
\ee
we project it to the AdS$_4$ bosonic spin-$s$ field $\phi_{\alpha_1 \dots \alpha_s, \dot{\alpha}_1 \dots \dot{\alpha}_s}(x)$ by 
\be
\Phi_{I_1 J_1, \dots, I_s J_s}(X) \mapsto \phi_{\alpha_1 \dots \alpha_s, \dot{\alpha}_1 \dots \dot{\alpha}_s}(x) = \left(\frac{{i}}{2L}\right)^sT_{(\alpha_1}^{I_1} \dots T_{\alpha_s)}^{I_s} \bar{T}_{(\dot{\alpha_1}}^{J_1} \dots \bar{T}_{\dot{\alpha}_s)}^{J_s} \Phi_{I_1 J_1, \dots, I_s J_s}(X(x)).
\ee
To make this a bijection we impose a homogeneity condition,
\be
\Phi_{I_1 J_1, \dots, I_s J_s}(\lambda X) = \lambda^w \Phi_{I_1 J_1, \dots, I_s J_s}(X),
\ee
and a transversality condition,  
\be \label{eq:transversality}
X^{I_1 J_1} \Phi_{I_1 J_1, \dots, I_s J_s}(X) =0.
\ee
This is equivalent to the usual approach to bosonic ambient tensors, as can be checked using \eqref{eq:projector}. For example, an ambient vector  $V^A(X)$ pulls back to an AdS vector $v^{\mu}(x)$,
\be \label{eq:pullback}
 V^A(X) \mapsto v_{\mu}(x)= \partial_{\mu} X^A(x) V_A(X(x)),
\ee
and this becomes a bijection by restricting to homogeneous ambient vectors satisfying the transversality condition $X_A V^A(X) = 0$.

\subsubsection*{Tensor-spinors}

We can describe tensor spinors by combining the previous two cases.  Let $s$ be a half-integer. Take an ambient space tensor $\Phi_{I_1 J_1, \dots, I_{s-\frac{1}{2}}  J_{s-\frac{1}{2}} , I}$ with the symmetries of the $J$-traceless tableau $[{s+\frac{1}{2}} ,{s-\frac{1}{2}} ]$,
\be
\Phi_{I_1 J_1, \dots, I_{{s-\frac{1}{2}} } J_{{s-\frac{1}{2}} }, I} \sim \raisebox{1.15ex}{\gyoung(_5{ {s+1/2} },_4{s-1/2} )} \, .
\ee
This ambient space tensor projects to a fermionic spin-$s$ AdS$_4$ field $\phi_{\alpha \alpha_1 \dots \alpha_{{s-\frac{1}{2}} }, \dot{\alpha}_1 \dots \dot{\alpha}_{{s-\frac{1}{2}} }}(x)$ by
\begin{align}
\Phi_{I_1 J_1, \dots, I_{{s-\frac{1}{2}} } J_{{s-\frac{1}{2}} }, I}(X) & \mapsto \phi_{\alpha \alpha_1 \dots \alpha_{{s-\frac{1}{2}} }, \dot{\alpha}_1 \dots \dot{\alpha}_{{s-\frac{1}{2}} }}(x) \nn \\
&= \frac{{i^{{s-\frac{1}{2}} }}}{(2L)^{s }}T_{(\alpha}^IT_{\alpha_1}^{I_1} \dots T_{\alpha_{{s-\frac{1}{2}} })}^{I_{s-\frac{1}{2}} } \bar{T}_{(\dot{\alpha_1}}^{J_1} \dots \bar{T}_{\dot{\alpha}_{s-\frac{1}{2}} )}^{J_{s-\frac{1}{2}} } \Phi_{I_1 J_1, \dots, I_{s-\frac{1}{2}}  J_{s-\frac{1}{2}} , I}(X(x)).
\end{align}
This map is a bijection if we restrict to ambient tensor-spinors that are homogeneous,
\be
\Phi^{I_1 J_1, \dots, I_{s-\frac{1}{2}}  J_{s-\frac{1}{2}} , I}(\lambda X) = \lambda^w \Phi^{I_1 J_1, \dots, I_{s-\frac{1}{2}}  J_{s-\frac{1}{2}} , I}(X) , 
\ee
and satisfy both a transversality condition and an eigenspinor equation,
\be
X_{I_1 J_1} \Phi^{I_1 J_1, \dots, I_{s-\frac{1}{2}}  J_{s-\frac{1}{2}} , I}(X) =0, \quad \mathcal{P}_{-}^I{}_J \Psi^{I_1 J_1, \dots, I_{s-\frac{1}{2}}  J_{s-\frac{1}{2}} ,J}(X) =0. 
\ee

\section{Shift symmetries of free fields on AdS$_4$}
\label{sec:AdS4-shifts}

In this section, we use the bispinor formalism described in Section \ref{sec:ambient} to explicitly write the shift symmetries of free fermion fields in AdS$_4$.

\subsection{Bosons}

We start by rewriting the known bosonic shift transformations in the bispinor language.  In the tensor language, a spin-$s$ boson $\phi_{\mu_1\cdots\mu_s}$ transforms under a level-$k$ shift as \cite{Bonifacio:2018zex}
\be
\delta \phi_{\mu_1\cdots\mu_s}= S_{A_1\cdots A_{s+k},B_1\cdots B_s}X^{A_1}\cdots X^{A_{s+k}} {\partial X^{B_1}\over \partial x^{\mu_1}}\cdots {\partial X^{B_s}\over \partial x^{\mu_s}} \, ,\label{bosonicshiftsymeetee}
\ee
where the $X^A$ are embedding coordinates and the tensor $S_{A_1\cdots A_{s+k},B_1\cdots B_s}$ is a fully traceless, constant ambient space tensor with permutation symmetries described by the Young tableau $[s+k,s]$.

In the spinor language, the bosonic spin-$s$ field has the index structure $\phi^{\alpha_1 \dots \alpha_s, \dot{\alpha}_1 \dots \dot{\alpha}_s}(x)$ and the shift symmetry \eqref{bosonicshiftsymeetee} takes the form
\begin{align} \label{eq:boson-shift-1}
\delta \phi^{\alpha_1 \dots \alpha_s, \dot{\alpha}_1 \dots \dot{\alpha}_s}(x) =& S^{ K_1 L_{1}, \dots K_k L_k,M_1N_1 \dots M_s N_s}\langle T_{K_1} T_{L_1} \rangle \dots \langle T_{K_k} T_{L_k} \rangle T^{\alpha_1}_{M_1} \bar{T}^{\dot{\alpha}_1}_{N_1} \dots T^{\alpha_s}_{M_s} \bar{T}^{\dot{\alpha}_s}_{N_s},
\end{align}
where $S^{ K_1 L_{1}, \dots, K_k L_k,M_1N_1 \dots M_s N_s}$ is a constant $J$-traceless ambient space tensor of type $[k+2s, k]$,
\be
S^{ K_1 L_{1}, \dots, K_k L_k,M_1N_1 \dots M_s N_s} \sim \raisebox{1.15ex}{\gyoung(_7{k+2s},_3k)},
\ee 
which is the $Sp(4, \mathbb{R})$ representation corresponding to the $SO(2,3)$ tensor representation $[s+k,s]$.
Using Eq.~\eqref{eq:cd} and ${\cal D}^2 \coloneqq {\cal D}_{\mu} {\cal D}^{\mu} = -\frac{1}{2} {\cal D}_{\alpha \dot{\alpha}} {\cal D}^{\alpha \dot{\alpha}}$, we can verify that the shift of the field is transverse and satisfies the Klein--Gordon equation,
\be
\left({\cal D}^2 -L^{-2}\left[s^2+2s(k+1)+k(k+3)\right] \right)\delta \phi^{ \alpha_1 \dots \alpha_s, \dot{\alpha}_1 \dots \dot{\alpha}_s} =0, \quad {\cal D}_{\alpha_1 \dot{\alpha}_1}  \phi^{ \alpha_1 \dots \alpha_s, \dot{\alpha}_1 \dots \dot{\alpha}_s} =0,
\ee
where the mass term is the correct one for a level-$k$ spin-$s$ shift field in $D=4$. 

On the corresponding ambient $[s,s]$-tensor $\Phi^{I_1 J_1, \dots, I_s J_s} $, the shift symmetry acts as
\begin{align} \label{eq:bosonic-spinor-shift}
\delta \Phi^{I_1 J_1, \dots, I_s J_s} & = S^{ K_1 L_{1}, \dots, K_k L_k,M_1N_1 \dots M_s N_s} X_{K_1 L_1}  \dots X_{K_k L_k}  X^{[I_1}{}_{M_{1}}  \delta^{J_1]}{}_{ N_1}  \dots X^{[I_s}{}_{M_s}  \delta^{J_s]}{}_{ N_s} ,
\end{align}
where we have assumed that the ambient field has weight $w= s+k$ (the transformation for other weights can be obtained by multiplying by an appropriate power of $\sqrt{-X^2}$) and we have redefined $S^{ K_1 L_{1}, \dots, K_k L_k,M_1N_1 \dots M_s N_s}$ by an overall factor compared to Eq.~\eqref{eq:boson-shift-1}.  
This transformation preserves the homogeneity and transversality constraints,
\begin{align}
\left( X^A\partial_A - s-k\right)  \delta\Phi_{I_1 J_1, \dots, I_s J_s} & = 0, \\
X^{I_1 J_1} \delta \Phi_{I_1 J_1, \dots, I_s J_s}& =0.
\end{align}

\subsection{Fermions}

It is now straightforward to extend the shift symmetries to fermions by adding another spinor index.  A fermionic spin-$s$ level-$k$ shift field $\phi^{\alpha \alpha_1 \dots \alpha_{s-\frac{1}{2}}, \dot{\alpha}_1 \dots \dot{\alpha}_{s-\frac{1}{2} }}$ has a shift symmetry that can be written as
\begin{align} 
\delta \phi^{ \alpha \alpha_1 \dots \alpha_{s-\frac{1}{2}}, \dot{\alpha}_1 \dots \dot{\alpha}_{s-\frac{1}{2}}}(x) = & S^{ K_1 L_{1}, \dots K_k L_k,M_1N_1 \dots M_{s-\frac{1}{2}} N_{s-\frac{1}{2}} I}\langle T_{K_1} T_{L_1} \rangle \dots \langle T_{K_k} T_{L_k} \rangle \nonumber \\
& \times T^{\alpha_1}_{M_1} \bar{T}^{\dot{\alpha}_1}_{N_1} \dots T^{\alpha_{s-\frac{1}{2}}}_{M_{s-\frac{1}{2}}} \bar{T}^{\dot{\alpha}_{s-\frac{1}{2}}}_{N_{s-\frac{1}{2}}}T_I^{\alpha}, \label{eq:fermion-shift-1}
\end{align}
where $S^{ K_1 L_{1}, \dots, K_k L_k,M_1N_1 \dots M_{s-\frac{1}{2}} N_{s-\frac{1}{2}} I}$ is a constant ambient space tensor with the permutation symmetries described by the $J$-traceless  tableau $[k+2s, k]$.
Using \eqref{eq:cd}, can verify that the shift of the field is transverse and satisfies the Weyl equation with the correct mass \eqref{shiftfermsymtmasse} in $D=4$, 
\begin{align}
& -i {\cal D}_{\alpha}{}^{(\dot{\alpha} |} \delta \phi^{\alpha \alpha_1 \dots \alpha_{s-\frac{1}{2}}, | \dot{\alpha}_1 \dots \dot{\alpha}_{s-\frac{1}{2}})} = \tilde{m} \delta \bar{\phi}^{\alpha_1 \dots \alpha_{s-\frac{1}{2}}, \dot{\alpha} \dot{\alpha}_1 \dots \dot{\alpha}_{s-\frac{1}{2}}}, \,\, \tilde{m} = (-1)^{s-\frac{1}{2}} L^{-1} \left(s+k+\frac{3}{2}\right), \\
& {\cal D}_{\alpha_1 \dot{\alpha}_1} \delta \phi^{\alpha \alpha_1 \dots \alpha_{s-\frac{1}{2}}, \dot{\alpha}_1 \dots \dot{\alpha}_{s-\frac{1}{2}}} =0\,.
\end{align}
This shows that the shift transformation is a symmetry of the free action.
On the corresponding ambient tensor $\Phi^{I_1 J_1, \dots, I_{s-\frac{1}{2}} J_{s-\frac{1}{2}}, I}$, the shift transformation is 
\begin{align}
\delta \Phi^{I_1 J_1, \dots, I_{s-\frac{1}{2}} J_{s-\frac{1}{2}}, I} & = \mathcal{P}_{+}^I{}_J S^{ K_1 L_{1}, \dots, K_k L_k,M_1N_1 \dots M_{s-\frac{1}{2}} N_{s-\frac{1}{2}} J}  X_{K_1 L_1}  \dots  X_{K_k L_k} \nn \\
&  \quad \times X^{[I_1}{}_{M_1}  \delta^{J_1]}{}_{ N_1}  \dots X^{[I_{s-\frac{1}{2}}}{}_{M_{s-\frac{1}{2}}}  \delta^{J_{s-\frac{1}{2}}]}{}_{ N_{s-\frac{1}{2}}}  ,
\end{align}
where we have taken the weight to be $w=s-\frac{1}{2}+k$ and we have redefined $S^{ K_1 L_{1}, \dots, K_k L_k,M_1N_1 \dots M_{s-\frac{1}{2}} N_{s-\frac{1}{2}} J}$ by an overall factor compared to Eq.~\eqref{eq:fermion-shift-1}.
This is consistent with the projection conditions,
\begin{align}
X^{I_1 J_1} \delta \Phi_{I_1 J_1, \dots, I_{s-\frac{1}{2}} J_{s-\frac{1}{2}}, I}& =0,\\
\mathcal{P}_{-}^J{}_I \delta\Psi^{I_1 J_1, \dots, I_{s-\frac{1}{2}} J_{s-\frac{1}{2}},I} & =0.
\end{align}

\section{Lie superalgebras}
\label{sec:superalgebras}

In this section, we search for Lie superalgebras to classify possible theories of AdS$_4$ fermions with nonlinearly realized symmetries.  Given a theory containing shift-symmetric fermions in AdS$_4$, the generators of isometries and the generators of the shift symmetries will form a Lie superalgebra containing the $\mathfrak{sp}(4, \mathbb{R})$ isometry algebra as a bosonic subalgebra.  The commutator between a shift generator and an $\mathfrak{sp}(4, \mathbb{R})$ generator is fixed by the fact that the shift generators all transform as some spinor or tensor representation of $\mathfrak{sp}(4, \mathbb{R})$.  This leaves only the (anti)commutators between the shift generators to be determined.  

In the free theory, the (anti)commutators between shift generators vanish, and the algebra trivially closes for any field content. 
We can always form trivial interactions by using powers of the field strengths \eqref{shiftinvfieldstrengthDde}, which are invariant under the linearized shifts.  These interactions do not modify the transformation laws or the algebra. Non-trivial interactions will modify the shift transformation laws by terms containing powers of the fields.  These modified transformation laws can then alter the commutators between shift generators so that they do not vanish.  We can thus search for possible non-trivial interacting theories by looking for suitable Lie superalgebras.

\subsection{Definition}

We first recall the definition of a Lie superalgebra (see Ref.~\cite{Kac1977} for an overview). A Lie superalgebra $\mathfrak{g}=\mathfrak{g}_0 \oplus \mathfrak{g}_1$ is a $\mathbb{Z}_2$-graded algebra with a Lie superbracket $[ \cdot, \cdot]: (\mathfrak{g}_i, \mathfrak{g}_j) \rightarrow \mathfrak{g}_{i+j \, {\rm mod} \, 2}$.  Elements of $\mathfrak{g}_0$ are called even elements and elements of $\mathfrak{g}_1$ are called odd elements.
Given $x\in \mathfrak{g}_i$, $y \in \mathfrak{g}_j$, $z \in \mathfrak{g}_k$, the Lie superbracket must satisfy the following two conditions:
\begin{enumerate}
\item The Lie superbracket of two odd elements is an anticommutator, and the Lie superbracket of two even elements, or of an even element and an odd element, is a commutator:
\be
[x, y] = -(-1)^{i j} [y, x] .
\ee
\item The super Jacobi identity:
\be
(-1)^{ik} [x, [y, z]] + (-1)^{i j} [ y, [z, x]] + (-1)^{jk} [ z, [x,y]] =0.
\ee
\end{enumerate}
In particular, these conditions imply that $\mathfrak{g}_0$ is an ordinary Lie algebra and that $\mathfrak{g}_1$ forms a representation of $\mathfrak{g}_0$ under the adjoint action.

One way to generate a Lie superalgebra is to start with an associative superalgebra $\mathfrak{g}=\mathfrak{g}_0 \oplus \mathfrak{g}_1$ and then define the Lie bracket in terms of the product of the associative algebra via 
\be \label{eq:bracket-from-product}
[x, y] \coloneqq  xy-(-1)^{ij} yx,
\ee
where $x\in \mathfrak{g}_i$ and $y \in \mathfrak{g}_j$. A particular example of this construction that we employ is to use the associative superalgebra of endomorphisms of a $\mathbb{Z}_2$-graded vector space $V = V_0 \oplus V_1$, where an endomorphism $\phi: V \rightarrow V$ is even if $\phi(V_i) \subset V_{i}$ and odd if $\phi(V_i) \subset V_{i+1\, {\rm mod} \, 2}$.\footnote{We are grateful to the anonymous referee of Ref.~\cite{Bonifacio:2019hrj} for pointing out the usefulness of this construction in the present context.}

\subsection{AdS superalgebras from endomorphisms}
Let $V = V_0 \oplus V_1$ be a graded vector space where $V_0$, the vector space of even elements, furnishes the trivial representation of $Sp(4, \mathbb{R})$ and $V_1$, the vector space of odd elements, furnishes the fundamental representation,
\be \label{eq:osp1|4-vector-space}
V = V_0 \oplus V_1= \left\{ \bullet \right\} \oplus \left\{ \gyoung(;) \right\}.
\ee
The set of endomorphisms of $V$ gives the associative superalgebra $\mathfrak{gl}(V)$, where the grading of an endomorphism is inherited from $V$ as defined above. Using the Lie superbracket defined in \eqref{eq:bracket-from-product},  we obtain a 
Lie superalgebra with generators transforming in the following $Sp(4, \mathbb{R})$ representations:
\be \label{eq:algebra1}
\left\{\bullet, \bullet, \gyoung(;;), \raisebox{1.15ex}{\gyoung(;,;)}  \right\} \oplus \left\{\gyoung(;), \gyoung(;)    \right\},
\ee
where the two sets correspond to the even and odd generators and we use $J_{IJ}$ to define an isomorphism between $V$ and $V^*$.
Restricting to the symmetric endomorphisms gives a Lie superalgebra with the generators
\be \label{eq:ospGenerators}
\left\{\bullet, \gyoung(;;)  \right\} \oplus \left\{ \gyoung(;)    \right\}.
\ee

\subsubsection{$\mathfrak{osp}(1|4)$}
Consider first the Lie superalgebra with generators given by \eqref{eq:ospGenerators}. We can derive the (anti)commutator relations from the (anti)commutators of $\mathfrak{gl}(V)$,
\be
\left[ M_A{}^B, M_C{}^D \right] = M_A{}^D \delta_C^B \pm M_C{}^B \delta_A^D.\label{glvcomree}
\ee
Denoting the symmetric tensor representation of $Sp(4, \mathbb{R})$ by $m_{IJ}$ and the fundamental by $Q_I$, we can break up $M_A{}^B$ as follows: with $A \in \{ I, 5\}$, we define $M_{AB} \coloneqq \epsilon_{BC} M_A{}^C$ with $\epsilon_{55} =1$, $\epsilon_{I5}=\epsilon_{5I}=0$, and $\epsilon_{IJ}= J_{IJ}$. Then for $M_{AB} = M_{BA}$ we take $m_{IJ} \coloneqq -2 M_{IJ}$ and $Q_I \coloneqq \alpha M_{I5}$ for an arbitrary constant $\alpha$, which is defined for later convenience in comparing to the field transformations.   From \eqref{glvcomree} we then get the following non-vanishing (anti)commutation relations: 
\begin{subequations} \label{eq:osp1|4}
\begin{align}
\left[ m_{IJ}, m_{KL} \right] & = - \left(m_{IL} J_{JK}+m_{IK} J_{JL}+m_{JL} J_{IK} +m_{JK} J_{IL} \right), \label{eq:isometry-commutator} \\
\left[ m_{IJ}, Q_{K} \right] & = - \left(Q_I J_{JK} +Q_J J_{IK} \right), \\
\left[ Q_{I}, Q_{J} \right] & = - \alpha^2 m_{IJ}.
\end{align}
\end{subequations}
The singlet $M_{55}$ commutes with everything and has therefore been dropped. The commutators \eqref{eq:isometry-commutator} are those of $\mathfrak{sp}(4, \mathbb{R})$.

This Lie superalgebra is familiar as the algebra of $\mathcal{N}=1$ supersymmetry in AdS$_4$, or equivalently the $\mathcal{N}=1$ superconformal algebra in three dimensions, namely $\mathfrak{osp}(1 | 4)$.  When the fermionic symmetry $Q_I$ is nonlinearly realized, this Lie superalgebra underlies the AdS$_4$ Akulov--Volkov theory \cite{Deser:1977uq, Zumino:1977av}, describing spontaneously broken supersymmetry, whose transformations we discuss in Section \ref{AVsection}. When $\alpha=0$, the symmetries are not modified from the free theory, as in the flat-space theory considered in Ref.~\cite{Roest:2017uga}.

In a similar fashion, we can get the algebra of higher-$\mathcal{N}$ SUSY in AdS$_4$, $\mathfrak{osp}(\mathcal{N} | 4)$, by starting with $\mathcal{N}$ singlets in $V_0$ in \eqref{eq:osp1|4-vector-space}.

\subsubsection{$\mathfrak{su}(2,2|1)$}

We can similarly find the (anti)commutation relations for the Lie superalgebra keeping all of the generators in \eqref{eq:algebra1}.
We again drop the decoupled trace, i.e., consider $\mathfrak{sl}(V)$, and we denote the generators by 
\be m_{IJ}, \ \ Q_I, \ \ s^{(2)}_{IJ}, \ \ s^{(1)}_I,\ \   s^{(0)},\label{ads5generatoree}\ee
where $m_{IJ}=m_{(IJ)}$ and $s^{(2)}_{IJ}=s^{(2)}_{[IJ]}$. 
 They are related to the components of $M_{AB}$ as follows: 
\begin{subequations}
\begin{align}
M_{IJ} &= -\frac{1}{2}m_{IJ} + \frac{1}{2} \lambda^{-1}s^{(2)}_{IJ}+\frac{1}{4} (\beta \sigma\lambda)^{-1} J_{IJ} s^{(0)}+J_{IJ} C,  \\
M_{I5} &=  \rho^{-1}\left( Q_I +\sigma^{-1} \lambda^{-1} s^{(1)}_I\right), \\
M_{5I} &=-\frac{1}{8}\rho \left( \alpha^{-1} Q_I - (\alpha \sigma \lambda)^{-1} s^{(1)}_I\right), \\
M_{55} &=(\beta \sigma \lambda)^{-1} s^{(0)}+C,
\end{align}
\end{subequations}
where $C$ is the generator we discard and $\rho$ is an arbitrary parameter that drops out of the algebra, reflecting the fact that a simultaneous rescaling of the fermionic generators can be absorbed into a redefinition of $\alpha$.

From \eqref{glvcomree} we get the following non-vanishing (anti)commutators:
\begin{subequations} \label{eq:su(2,2|1)-B}
\begin{align}
\left[ m_{IJ}, m_{KL} \right]& = - \left( J_{IK} m_{JL} + J_{IL} m_{JK} + J_{JL} m_{IK} + J_{JK} m_{IL} \right), \\
\left[m_{IJ}, Q_K \right] &= - \left( J_{IK} Q_J +J_{JK} Q_I \right) , \\
\left[ m_{IJ}, s^{(2)}_{KL} \right] & = -\left(J_{IK} s^{(2)}_{JL}-J_{IL} s^{(2)}_{JK}-J_{JL} s^{(2)}_{IK}+ J_{JK} s^{(2)}_{IL}\right), \\
\left[m_{IJ}, s^{(1)}_K \right] & = -\left( J_{IK} s^{(1)}_J + J_{JK} s^{(1)}_I \right), \\
\left[Q_I, Q_J \right] & = 2 \alpha m_{IJ}, \\
\left[ Q_I, s^{(2)}_{JK} \right] & = - \sigma^{-1}  \left( J_{IJ} s^{(1)}_K -J_{IK} s^{(1)}_J + \frac{1}{2} J_{JK} s^{(1)}_I \right), \\
\left[ Q_I, s^{(1)}_J \right] & = 2\sigma \alpha  s_{IJ}^{(2)} -3  \alpha \beta^{-1} J_{IJ} s^{(0)}, \\
\left[ Q_I , s^{(0)} \right] & =\beta  s^{(1)}_I, \\ 
\left[ s^{(2)}_{IJ}, s^{(2)}_{KL} \right] & = \lambda^2 \left( J_{IK} m_{JL} - J_{IL} m_{JK} + J_{JL} m_{IK} - J_{JK} m_{IL} \right), \\
\left[ s^{(2)}_{IJ}, s^{(1)}_K \right] & = \sigma \lambda^2 \left( J_{JK} Q^{(1)}_I -J_{IK} Q^{(1)}_J + \frac{1}{2} J_{IJ} Q^{(1)}_K \right), \\
\left[s^{(1)}_I, s^{(1)}_J \right] & = -2\alpha \sigma^2 \lambda^2 m_{IJ}, \\
\left[s^{(1)}_I, s^{(0)} \right] & = \beta \sigma^2 \lambda^2 Q^{(1)}_I \, .
\end{align}
\end{subequations}

This Lie superalgebra is the algebra of $\mathcal{N}=1$ supersymmetry in AdS$_5$, i.e., the $\mathcal{N}=1$ superconformal algebra in four dimensions, namely $\mathfrak{su}(2,2|1)$.  The bosonic generators $m_{IJ},  \ s^{(2)}_{IJ}$ form an $\mathfrak{so}(4,2)$ subalgebra, the algebra of isometries of AdS$_5$, and the remaining singlet $s^{(0)}$ corresponds to the generator of the $\mathfrak{u}(1)$ $R$-symmetry in AdS$_5$.  The bosonic subalgebra is thus $\mathfrak{so}(2,4) \times \mathfrak{u}(1)$.
The generators $m_{IJ}, \ Q_I$ form an $\mathfrak{osp}(1|4)$ subalgebra.  The bosonic part $m_{IJ}$ is the AdS$_4$ isometry algebra $\mathfrak{so}(2,3)$. 

We can nonlinearly realize this algebra with the subalgebra $\mathfrak{osp}(1 | 4)$ linearly realized, corresponding to the breaking pattern
\be
\mathfrak{su}(2,2|1) \rightarrow \mathfrak{osp}(1|4) \,.
\ee
This corresponds to the spontaneous symmetry breaking of the minimal five-dimensional AdS SUSY to the minimal four-dimensional AdS SUSY. This symmetry breaking pattern underlies the supersymmetric DBI theory on AdS$_4$ \cite{Ivanov:1979ft, Love:2006hv}.\footnote{The bosonic part of this breaking pattern is
\be
\mathfrak{so}(2,4) \times \mathfrak{u}(1) \rightarrow \mathfrak{so}(3,2) .
\ee
This corresponds to the DBI theory describing an AdS$_4$ brane embedded into an AdS$_5$ bulk \cite{Clark:2005ht, Goon:2011qf,Goon:2011uw}, along with an additional $u(1)$ shift-symmetric scalar (see \cite{Hinterbichler:2012fr} for a physical realization of this $u(1)$).  
} The generators $s^{(2)}_{IJ}$, $s^{(1)}_I$, and $s^{(0)}$ correspond to non-linearly realized shift symmetries for a shift-symmetric $k=1$ scalar, a $k=1$ spin-1/2 fermion, and a $k=0$ scalar, respectively. These fields form an ${\cal N}=1$ linear SUSY multiplet on AdS$_4$. See Ref.~\cite{Clark:2005ma} for the coset construction of a theory realizing the breaking of $\mathfrak{su}(2,2|1)$ to the 4D Poincar\'e algebra and Ref.~\cite{Delduc:2001tb} for $OSp(4,1)$ broken to 3d $\mathcal{N}=1$ Poincar\'e SUSY.

\subsubsection{SUSY multiplets}
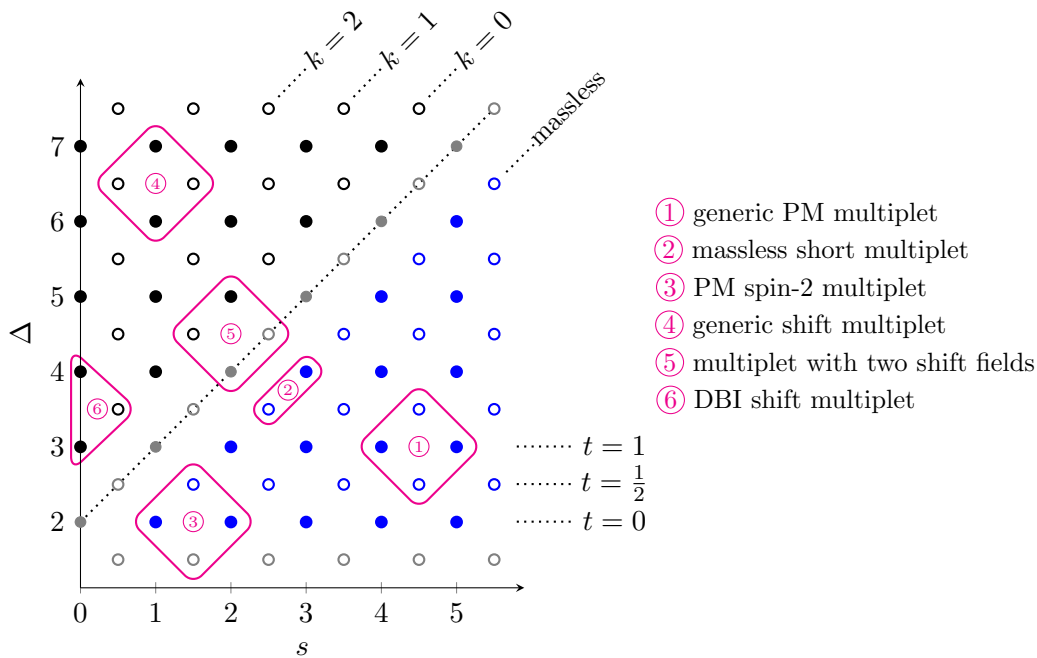
\begin{figure}[ht!]
\begin{center}
\begin{tikzpicture}
\node at (7.1, 0.89){$t=0$};
\draw [thick, dotted] (5.8,0.88) -- (6.54 ,0.88);
\node at (7.1, 1.39){$t=\frac{1}{2}$};
\draw [thick, dotted] (5.8,1.38) -- (6.54 ,1.38);
\node at (7.1, 1.89){$t=1$};
\draw [thick, dotted] (5.8,1.88) -- (6.54 ,1.88);
 \node[rotate=45] at (5.35,7.3) {$k=0$};
\draw [thick, dotted] (4.63,6.53) -- (5,6.91);
 \node[rotate=45] at (4.35,7.3) {$k=1$};
\draw [thick, dotted] (3.63,6.53) -- (4,6.91);
 \node[rotate=45] at (3.35,7.3) {$k=2$};
\draw [thick, dotted] (2.63,6.53) -- (3,6.91);
 \node[rotate=45] at (6.46,6.4) {{\small massless}};
\draw [thick, dotted] (5.66,5.53) -- (6.03,5.9);
 \node at (4.5,1.88) {\color{magenta}{\tiny \circled{1}}};
 \node at (2.76,2.63) {{\color{magenta} \tiny \circled{2}}};
 \node at (1.5,0.88) {{\color{magenta}\tiny \circled{3}}};
 \node at (1,5.38) {{\color{magenta}\tiny \circled{4}}};
 \node at (2,3.38) {{\color{magenta}\tiny \circled{5}}};
 \node at (0.225,2.38) {{\color{magenta}\tiny \circled{6}}};
 \node at (9.528,5) {\small {\color{magenta} \circled{1}} generic PM multiplet};
 \node at (9.723,4.5) {{\small {\color{magenta}\circled{2}} massless short multiplet}};
 \node at (9.455,4) {{\small {\color{magenta}\circled{3}} PM spin-2 multiplet}};
 \node at (9.578,3.5) {{\small {\color{magenta} \circled{4}} generic shift multiplet}};
 \node at (10.17,3) {{\small {\color{magenta}\circled{5}} multiplet with two shift fields}};
 \node at (9.38,2.5) {{\small {\color{magenta} \circled{6}} DBI shift multiplet}};
%\draw[rounded corners, thick, magenta] (1, -1) rectangle (2, 0) {};
\node[draw, thick, magenta, diamond, scale=3.3, rounded corners=4.5pt] at (1.5, .88) {};
\node[draw, thick, magenta, diamond, scale=3.3, rounded corners=4.5pt] at (2, 3.38) {};
\node[draw, thick, magenta, diamond, scale=3.3, rounded corners=4.5pt] at (1, 5.38) {};
\node[draw, thick, magenta, diamond, scale=3.3, rounded corners=4.5pt] at (4.5, 1.88) {};
    \draw [thick, magenta, rounded corners=5.5pt] (-.125,1.55)--(.75,2.38)--(-.125,3.18)--cycle;
    \draw [thick, magenta, rounded corners=4pt] (2.25, 2.37)--(2.5,2.12)--(3.26,2.88)--(3.01, 3.13)--cycle;
	\begin{axis}[%
	axis y line=middle, axis x line=bottom, x=1cm, y=1cm,
	xlabel=$s$, xlabel near ticks,
	ylabel= $\Delta$, ylabel near ticks,
	xmin=0, xmax = 5.9, ymax=6.9, ymin=1/8,
	yticklabels={$2$, $3$, $4$, $5$, $6$, $7$},ytick={1,...,6},
	xtick={0, ...,7},
	scatter/classes={%
		a={mark=*, thick, blue},%
		b={mark=o, thick, blue},%
		c={mark=o, thick, black},%
		d={mark=*, thick, black},%
		e={mark=o, thick, gray},%
		f={mark=*, gray, mark options={scale=2, fill=white}},%
		w={mark=*, scale=0.5, white}%
		}]
\addplot[black, thick, dotted, domain=0:5.5] {x+1};
	\addplot[scatter, only marks,%
		scatter src=explicit symbolic]%
	table[meta=label] {
x     y      label
1   1   a 
2   1   a 
3   1   a 
4   1   a 
5   1   a 
6   1   a 
7   1   a 
2   2   a 
3   2   a 
4   2   a 
5   2   a 
6   2   a 
7   2   a 
3   3   a 
4   3   a 
5   3   a 
6   3   a 
7   3   a 
4   4   a 
5   4   a 
6   4   a 
7   4   a 
5   5   a 
6   5   a 
7   5   a 
6   6   a 
7   6   a 
7 7 a
1.5 1.5   b
2.5 1.5   b
3.5 1.5   b
4.5 1.5   b
5.5 1.5   b
6.5 1.5   b
2.5 2.5   b
3.5 2.5   b
4.5 2.5   b
5.5 2.5   b
6.5 2.5   b
3.5 3.5   b
4.5 3.5   b
5.5 3.5   b
6.5 3.5   b
4.5 4.5   b
5.5 4.5   b
6.5 4.5   b
5.5 5.5   b
6.5 5.5   b
6.5 6.5   b
0 2 d
0 3 d
0 4 d
0 5 d
0 6 d
1 3 d
1 4 d
1 5 d
1 6 d
2 4 d
2 5 d
2 6 d
3 5 d
3 6 d
4 6 d
.5 2.5 c
.5 3.5 c
.5 4.5 c
.5 5.5 c
.5 6.5 c
1.5 3.5 c
1.5 4.5 c
1.5 5.5 c
1.5 6.5 c
2.5 4.5 c
2.5 5.5 c 
2.5 6.5 c
3.5 5.5 c
3.5 6.5 c
4.5 6.5 c
.5 .5 e
1.5 .5 e
2.5 .5 e
3.5 .5 e
4.5 .5 e
5.5 .5 e
0 1 f
1 2 f
2 3 f
3 4 f
4 5 f
5 6 f
.5 1.5 e
1.5 2.5 e
2.5 3.5 e
3.5 4.5 e
4.5 5.5 e
5.5 6.5 e
.5 1.5 w
1.5 2.5 w
2.5 3.5 w
3.5 4.5 w
4.5 5.5 w
5.5 6.5 w
	};
	\end{axis}
\end{tikzpicture}
\caption{${\cal N}=1$ SUSY multiplets of PM and shift-symmetric fields on AdS$_4$.  Filled circles are bosons and unfilled circles are fermions, blue circles are PM fields, black circles are shift-symmetric fields, and grey circles are neither PM nor shift symmetric but participate in some of the multiplets with the PM or shift-symmetric fields. The supermultiplets labelled {\color{magenta}\textcircled{\raisebox{-0.9pt}{4}}}, {\color{magenta}\textcircled{\raisebox{-0.9pt}{5}}}, and {\color{magenta}\textcircled{\raisebox{-0.9pt}{6}}} are the longitudinal modes of the supermultiplets labelled {\color{magenta} \textcircled{\raisebox{-0.9pt}{1}}}, {\color{magenta}\textcircled{\raisebox{-0.9pt}{2}}}, and {\color{magenta} \textcircled{\raisebox{-0.9pt}{3}}}, respectively.}
\label{fig:fermion-susy}
\end{center}
\end{figure}

Generic massive ${\cal N}=1$ SUSY multiplets on AdS$_4$ consist of four states that form a diamond in the $(s, \Delta)$-plane \cite{Heidenreich:1982rz}. The states $D(\Delta,s), \ D(\Delta+{1\over 2} ,s+{1\over 2} ),\ D(\Delta+{1\over 2},s-{1\over 2} ),\ D(\Delta+1,s)$ form an ${\cal N}=1$ SUSY multiplet, where the first state is a primary which labels the multiplet and the remaining states are filled out by applying the fermionic operators. The generic massive non-unitary multiplets also take this form \cite{Garcia-Saenz:2018wnw, Buchbinder:2019olk, Bittermann:2020xkl}. In the region where there are PM fields, this diamond can overlay precisely onto PM points, so the PM fields form supermultiplets with each other.  These supermultiplets shorten at the top boundary of this region, where the fields become massless, so that the massless fields form short supermultiplets containing only two states.  See Figure \ref{fig:fermion-susy}.

In the region where there are shift-symmetric fields, the diamond can also overlay precisely onto the shift-symmetric points, so the shift-symmetric fields also form supermultiplets with each other, as shown in Figure \ref{fig:fermion-susy}.  At the left boundary of the region of shift-symmetric fields where $s=0$, the supermultiplet diamond degenerates to three states since the spin cannot be less than zero.  The multiplet with SUSY primary $D(3,0)$ thus contains the states $D(3,0), \ D({7\over 2} ,{1\over 2} ),\ D(4,0)$, which is precisely the field content of the SUSY DBI theory discussed above.  

A generic massive graviton supermultiplet, i.e, a supermultiplet whose highest-spin state is a massive spin-2 field, has the states  $D(\Delta,{3\over 2}), \ D(\Delta+{1\over 2} ,2 ),\ D(\Delta+{1\over 2},1 ),\ D(\Delta+1,{3\over 2})$.  In the limit $\Delta\rightarrow {3\over 2}$ this becomes a partially massless graviton multiplet.  The longitudinal modes that decouple are precisely those of the SUSY DBI theory: the $\Delta=4$ scalar is the longitudinal mode of the massive spin-2 that decouples as it becomes partially massless, the $\Delta=3$ scalar is the longitudinal mode of the massive spin-1 field that decouples as it becomes massless, and the $\Delta={7\over 2}$ spin-$1/2$ fermion is the longitudinal mode of the massive spin-$3/2$ field that decouples as it becomes massless (this branching rule is described in Eq. 4.4 of Ref.~\cite{Garcia-Saenz:2018wnw}).  

\subsection{Lie superalgebras for extended fermionic shift symmetries}

It follows from the discussion in Section~\ref{sec:ambient} that the generator of a level-$k$ shift symmetry on a spin-$s$ field is described by the $\mathfrak{sp}(4, \mathbb{R})$ representation described by the Young tableau $[k+2s, k]$. In the above two examples of Lie superalgebras, the fermionic generators transformed in the fundamental representation of $\mathfrak{sp}(4, \mathbb{R})$, which therefore correspond to $k=0$ shift symmetries when nonlinearly realized on a spin-$\frac{1}{2}$ field. We can find more complicated Lie superalgebras that can underlie theories of spin-$\frac{1}{2}$ fields with $k>0$ shift symmetries by generalizing the above construction to find Lie superalgebras containing an $\mathfrak{sp}(4, \mathbb{R})$ subalgebra and including generators of the type $[k+1, k]$. 

As an example, take the $\mathbb{Z}_2$-graded vector space made from the following $\mathfrak{sp}(4, \mathbb{R})$ representations:
\be
 V=\left\{ \bullet \right\} \oplus \left\{ \raisebox{1.15ex}{\gyoung(;;,;)} \right\}.
\ee
Taking the symmetric part of the endormorphisms of $V$ and dropping the overall trace, we get a Lie superalgebra with the generators
\be \label{eq:algebra2}
\left\{\gyoung(;;), \gyoung(;;), \raisebox{1.15ex}{\gyoung(;;;,;)}, \raisebox{1.15ex}{\gyoung(;;;;,;;)} \right\} \oplus \left\{\raisebox{1.15ex}{\gyoung(;;,;)}  \right\}.
\ee
Interpreted as the symmetry algebra of shift-symmetric fields, this would correspond to a theory of a $k=1$ spin-$\frac{1}{2}$ fermion interacting with three spin-1 fields with $k=0$, $k=1$, and $k=2$ shift symmetries. This Lie superalgebra is $\mathfrak{osp}(1 | 16)$ and is a candidate algebra for a $k=1$ analogue of the Akulov--Volkov theory.\footnote{We checked by writing down the most general (anti)commutators and trying to satisfy the super Jacobi identities that there is no Lie superalgebra with just the generators $\left\{\gyoung(;;)\right\} \oplus \left\{\raisebox{1.15ex}{\gyoung(;;,;)}  \right\}$ and a non-vanishing commutator between shifts, so additional fields beyond the $k=1$ spin 1/2 fermion are necessary. We can also check that among the classical simple Lie superalgebras there are none with the correct number of bosonic and fermionic generators.  This implies there is no DBI-like theory for a spin-1/2 fermion on AdS$_4$ without additional fields.}

The generalization of this construction to spin-$\frac{1}{2}$ fermions with higher values of $k$ is to consider the endomorphisms of the $\mathbb{Z}_2$-graded vector space
\be
V^{(k)}=\left\{ \bullet \right\} \oplus \left\{ \raisebox{1.15ex}{\gyoung(_5{k+1},_4k)} \right\}.
\ee
Taking the symmetric part and dropping the overall trace gives a Lie superalgebra with generators
\be
\left\{ \rm{Sym}^2\left( \raisebox{1.15ex}{\gyoung(_5{k+1},_4k)} \right) \right\} \oplus \left\{ \raisebox{1.15ex}{\gyoung(_5{k+1},_4k)} \right\},
\ee
where $\rm{Sym}^2$ denotes the symmetric square, e.g., $\rm{Sym}^2\left( \gyoung(;)   \right)=  \gyoung(;;) $.
The representations of the bosonic generators can be found recursively from
\begin{align}
{ \rm Sym}^2\left( [k+2, k+1] \right)& = {\rm Sym}^2\left( [k+1, k] \right)\oplus \bigoplus_{j=1}^{\lfloor \frac{k+1}{2} \rfloor} \left[2k+3, 2(k-2j)+3 \right] \nonumber \\
&\oplus \bigoplus_{l=0}^{2}\bigoplus_{j=1}^{\lfloor \frac{k+2+\lfloor l/2 \rfloor}{2} \rfloor} \left[2(k+1)+l, 2(k-2j+2)+l \right],
\end{align}
where we checked this formula up to $k=79$ using the program \texttt{LiE}. When using these algebras  as nonlinear shift symmetries, this means that having a spin-$\frac{1}{2}$ fermion with a level $k$ shift with $k>0$  requires bosonic particles with spin up to $2 \lfloor k/2 \rfloor+1$.
These Lie superalgebras are $\mathfrak{osp}\left(1 |n_k \right)$, where
\be
n_k =  \frac{2}{3}(k+1)(k+2)(k+3)
\ee
is the dimension of the representation $[k+1, k]$.
These are analogous to the finite-dimensional higher-spin  bosonic algebras considered in Ref.~\cite{Joung:2015jza}. 

\subsection{No supersymmetric special galileon}

We now comment on whether there exists a Lie superalgebra that could underlie a supersymmetric version of the special galileon in AdS$_4$. The special galileon in AdS$_4$ is a theory of a single $k=2$ shift-symmetric scalar that has non-trivial interactions realizing the symmetry breaking pattern $\mathfrak{sl}(5, \mathbb{R}) \rightarrow \mathfrak{so}(3,2)$ \cite{Bonifacio:2018zex,Bonifacio:2021mrf}.  It is known that there is no supersymmetric special galileon in flat spacetime \cite{Elvang:2018dco, Roest:2019dxy}, but this does not necessarily rule out the existence of such a theory for nonzero curvature since the flat limit could be trivial or singular. 

Looking at Figure~\ref{fig:fermion-susy}, the simplest $\mathcal{N}=1$ multiplet containing the special galileon would be the one shifted upwards by one unit of $\Delta$ from the SUSY DBI multiplet, i.e., the multiplet with a $k=1$ scalar and a $k=1$ fermion in addition to the $k=2$ special galileon. Altogether, the generators of the shift symmetries of the fields in this multiplet belong to the following $\mathfrak{sp}(4, \mathbb{R})$ representations:
\be \label{eq:algebra2}
\left\{\raisebox{1.15ex}{\gyoung(;,;)},\raisebox{1.15ex}{\gyoung(;;,;)} , \raisebox{1.15ex}{\gyoung(;;,;;)}  \right\}.
\ee
We would therefore like to find a Lie superalgebra containing these generators and the generators of the ${\cal N}=1$ algebra $\mathfrak{osp}(1|4)$.
A direct search (i.e., writing all possible structures for the commutators and demanding that they satisfy the super Jacobi identities) shows that no Lie superalgebra exists with just these generators and a nonvanishing commutator between the shifts. We can also check that among the classical simple Lie superalgebras there are none with the correct number of bosonic and fermionic generators. This shows that there is no supersymmetric special galileon in AdS$_4$ without additional SUSY multiplets.\footnote{We could attempt to find an appropriate enlarged algebra by adding additional generators, corresponding to additional Goldstone modes. One approach that does \textit{not} work is considering the endormorphisms of the graded vector space
\be
V = \left\{ \bullet, \raisebox{1.15ex}{\gyoung(;,;)} \right\} \oplus \left\{ \gyoung(;) \right\}.
\ee
Taking the symmetric part, this will have both $\mathfrak{sl}(5, \mathbb{R})$ and $\mathfrak{osp}(1|4)$ subalgebras, but the special galileon shifts are scalars under the isometry subalgebra.}

\section{Akulov--Volkov transformations}
\label{AVsection}

We now discuss an example of nonlinear realizations of the above Lie superalgebras on fields, the Akulov--Volkov theory. This theory nonlinearly realizes the Lie superalgebra $\mathfrak{osp}(1|4)$  on a single $k=0$ spin-1/2 field, the goldstino. We use the ambient space formalism to write the transformations in a way that is manifestly covariant under $Sp(4,\mathbb{R})$ transformations.

The action of the isometries on a fermion using $Sp(4,\mathbb{R})$ generators is given by
\be
\delta_{m_{IJ}} \Psi_K = 2 \left( X^L{}_I \partial_{LJ} + X^L{}_J \partial_{LI}\right) \Psi_K+J_{IK} \Psi_J+J_{JK} \Psi_I.
\ee
This transformation is consistent with the projection constraints, 
\be
\mathcal{P}_{\pm}^L{}_K \Psi^K =0 \implies \mathcal{P}_{\pm}^L{}_K  \delta_{m_{IJ}} \Psi^K =0,
\ee
and satisfies the $\mathfrak{sp}(4, \mathbb{R})$ commutation relations,
\be
 \left[\delta_{m_{IJ}}, \delta_{m_{KL}} \right]  \Psi_M  =  \left( J_{JK} \delta_{m_{IL}} + J_{JL} \delta_{m_{IK}}+J_{IK} \delta_{m_{JL}} +J_{IL} \delta_{m_{JK}}  \right)\Psi_M .
\ee

We can write the field-independent part of the $k=0$ shift, i.e., the broken SUSY generator, on a spin-$\frac{1}{2}$ fermion as
\be
\delta_{Q_{I}} \Psi^K = \mathcal{P}_{\pm}^K{}_I, \quad  \mathcal{P}_{\mp }^L{}_K \Psi^K =0,
\ee
i.e., $\delta_{Q_{I}} \Psi^K = \mathcal{P}_{+}^K{}_I$ for a left-handed fermion and $\delta_{Q_{I}} \Psi^K = \mathcal{P}_{-}^K{}_I$ for a right-handed fermion.

We now need to find the field-dependent parts of the shift transformation, which are due to interactions deforming the algebra of the free theory.  Consider the following ansatz for the full non-linear action of the shift generator $Q_I$:
\begin{align} 
\delta_{Q_{I}} \Psi^K = \mathcal{P}_{\pm}^K{}_I + \beta \Psi_I \Psi^K + \gamma \Psi^M X^{L}{}_M \partial_{LI} \Psi^K + \gamma \Psi^M X^{L}{}_{I} \partial_{LM} \Psi^K + \frac{1}{2} \gamma \Psi_M \Psi^M  \mathcal{P}_{\mp}^K{}_I,\label{avtraanse}
\end{align}
where $\beta$ and $\gamma$ are coefficients to be determined.
The structure of \eqref{avtraanse} has been chosen so that the ansatz projects correctly for any choice of the coefficients,
\be
\mathcal{P}_{\mp }^L{}_K \Psi^K =0 \implies \mathcal{P}_{\mp}^L{}_K  \delta_{Q_{I}} \Psi^K =0.
\ee
The ansatz \eqref{avtraanse} transforms as a spinor under the isometries, so we have the commutator
\be
[\delta_{m_{IJ}}, \delta_{Q_K}] \Psi^L =  J_{JK} \delta_{Q_I} \Psi^L +J_{IK} \delta_{Q_J} \Psi^L.
\ee
If we set $\beta= \gamma/2=\alpha^2$ and compute the anti-commutator of the broken supersymmetry transformation, then a long calculation with many nontrivial cancellations gives
\be
[ \delta_{Q_I}, \delta_{Q_J} ]\Psi_K = \alpha^2 \delta_{m_{IJ}} \Psi_K,
\ee
so with these choices of coefficients the transformation \eqref{avtraanse} gives a nonlinear realization of $\mathcal{N}=1$ supersymmetry in AdS$_4$.

To summarize, we obtain the following covariant form of the nonlinear action of the $\mathcal{N}=1$ AdS$_4$ supersymmetry algebra $\mathfrak{osp}(1|4)$ on a goldstino:
\begin{align}
\delta_{m_{IJ}} \Psi^K & = -4X^L{}_{(I} \partial_{J)L}  \Psi^K-2\delta^K{}_{(I} \Psi_{J)}, \\
\delta_{Q_{I}} \Psi^K & = \mathcal{P}_{\pm}^K{}_I +  \alpha^2\left[ \Psi_I \Psi^K - 4  \Psi^M X^{L}{}_{(M} \partial_{I)L} \Psi^K + \Psi_M \Psi^M  \mathcal{P}_{\mp}^K{}_I \right],
\end{align}
where $\mathcal{P}_{\mp }^L{}_K \Psi^K=0$.

\section{Conclusions}
\label{sec:conclusions}

In this paper, we have classified the extended shift symmetries acting on free fermionic fields on (A)dS space. For a given spin, these shifts occur at specific mass values relative to the background (A)dS curvature scale and can be labelled by a non-negative integer $k$, where $k=0$ corresponds to a constant shift, $k=1$ is a shift linear in the co-ordinates, $k=2$ is a shift quadratic in the coordinates, and so on.  This completes the classification of the possible shift symmetries acting on free covariant fields, with some exotic exceptions such as continuous spin particles and anyons in lower dimensions.  In addition, for spacetime dimension $D=4$ we have studied algebras that could underly non-trivial interactions of these fermions, giving a method for constructing an infinite number of Lie superalgebras that can underly theories that include shift-symmetric fermions.
The simplest such algebra is the $\mathcal{N}=1$ AdS$_4$ SUSY algebra, which when broken to the AdS$_4$ isometries underlies the AdS version of the Akulov--Volkov model governing the goldstino mode of spontaneous SUSY breaking.  Another, more complicated, algebra gives the supersymmetric DBI model describing the breaking of the AdS$_5$ SUSY algebra down to the AdS$_4$ SUSY algebra.  We have also used the classification of Lie superalgebras to rule out a possible SUSY version of the special galileon on AdS$_4$.

An interesting future direction would be to investigate the explicit interacting Lagrangians using the formalism we describe, as for the bosonic examples given in Refs.~\cite{Bonifacio:2018zex,DeRham:2018axr,Bonifacio:2021mrf}. The coset construction \cite{Coleman:1969sm,Callan:1969sn,Zumino:1970tu,Volkov:1973vd} accomplishes this in principle, as in Refs.~\cite{Zumino:1977av, Ivanov:1979ft, Delduc:2001tb, Love:2006hv, Clark:2005ma}, but in practice it can become unwieldy and intractable to extract explicit interactions and transformations.

Note that the AdS galileon interactions \cite{Goon:2011qf,Goon:2011uw,Burrage:2011bt} do not deform the symmetry algebra of the free $k=1$ scalar theory \cite{Bonifacio:2018zex,Bonifacio:2021mrf}. The interactions are instead singled out because they do not generate higher-order equations of motion.
This is different from the status of the flat-space galileons \cite{Nicolis:2008in}, which are Wess-Zumino terms \cite{Goon:2012dy}, i.e., terms which do not deform the symmetry but cannot be written in terms of the invariant field strengths.    It would also be interesting to classify such interactions for the fermions in (A)dS space.

\paragraph{Acknowledgements:} We would like to thank Diederik Roest and David Stefanyszyn for helpful discussions. K.H. acknowledges support from DOE grant DE-SC0009946.

%\appendix

%\newpage
\renewcommand{\em}{}
\bibliographystyle{utphys}
\addcontentsline{toc}{section}{References}
\bibliography{fermion-arxiv}

\providecommand{\href}[2]{#2}\begingroup\raggedright\begin{thebibliography}{100}

\bibitem{Luty:2003vm}
M.~A. Luty, M.~Porrati, and R.~Rattazzi, ``{Strong interactions and stability
  in the DGP model},''
  \href{http://dx.doi.org/10.1088/1126-6708/2003/09/029}{{\em JHEP} {\bf 09}
  (2003)  029}, \href{http://arxiv.org/abs/hep-th/0303116}{{\tt
  arXiv:hep-th/0303116}}.

\bibitem{Nicolis:2008in}
A.~Nicolis, R.~Rattazzi, and E.~Trincherini, ``{The Galileon as a local
  modification of gravity},''
  \href{http://dx.doi.org/10.1103/PhysRevD.79.064036}{{\em Phys. Rev. D} {\bf
  79} (2009)  064036}, \href{http://arxiv.org/abs/0811.2197}{{\tt
  arXiv:0811.2197 [hep-th]}}.

\bibitem{Cheung:2014dqa}
C.~Cheung, K.~Kampf, J.~Novotny, and J.~Trnka, ``{Effective Field Theories from
  Soft Limits of Scattering Amplitudes},''
  \href{http://dx.doi.org/10.1103/PhysRevLett.114.221602}{{\em Phys. Rev.
  Lett.} {\bf 114} (2015) no.~22, 221602},
  \href{http://arxiv.org/abs/1412.4095}{{\tt arXiv:1412.4095 [hep-th]}}.

\bibitem{Hinterbichler:2015pqa}
K.~Hinterbichler and A.~Joyce, ``{Hidden symmetry of the Galileon},''
  \href{http://dx.doi.org/10.1103/PhysRevD.92.023503}{{\em Phys. Rev. D} {\bf
  92} (2015) no.~2, 023503}, \href{http://arxiv.org/abs/1501.07600}{{\tt
  arXiv:1501.07600 [hep-th]}}.

\bibitem{Cheung:2016drk}
C.~Cheung, K.~Kampf, J.~Novotny, C.-H. Shen, and J.~Trnka, ``{A Periodic Table
  of Effective Field Theories},''
  \href{http://dx.doi.org/10.1007/JHEP02(2017)020}{{\em JHEP} {\bf 02} (2017)
  020}, \href{http://arxiv.org/abs/1611.03137}{{\tt arXiv:1611.03137
  [hep-th]}}.

\bibitem{Novotny:2016jkh}
J.~Novotny, ``{Geometry of special Galileons},''
  \href{http://dx.doi.org/10.1103/PhysRevD.95.065019}{{\em Phys. Rev. D} {\bf
  95} (2017) no.~6, 065019}, \href{http://arxiv.org/abs/1612.01738}{{\tt
  arXiv:1612.01738 [hep-th]}}.

\bibitem{vilenkin1978special}
N.~Vilenkin, {\em Special Functions and the Theory of Group Representations}.
\newblock Translations of mathematical monographs. American Mathematical Soc.,
  1978.
\newblock \url{https://books.google.com/books?id=08hPoGgSQFIC}.

\bibitem{Allen:1985ux}
B.~Allen, ``{Vacuum States in de Sitter Space},''
  \href{http://dx.doi.org/10.1103/PhysRevD.32.3136}{{\em Phys. Rev. D} {\bf 32}
  (1985)  3136}.

\bibitem{Allen:1987tz}
B.~Allen and A.~Folacci, ``{The Massless Minimally Coupled Scalar Field in De
  Sitter Space},'' \href{http://dx.doi.org/10.1103/PhysRevD.35.3771}{{\em Phys.
  Rev. D} {\bf 35} (1987)  3771}.

\bibitem{Antoniadis:1991fa}
I.~Antoniadis and E.~Mottola, ``{4-D quantum gravity in the conformal
  sector},'' \href{http://dx.doi.org/10.1103/PhysRevD.45.2013}{{\em Phys. Rev.
  D} {\bf 45} (1992)  2013--2025}.

\bibitem{Folacci:1992xc}
A.~Folacci, ``{BRST quantization of the massless minimally coupled scalar field
  in de Sitter space: Zero modes, euclideanization and quantization},''
  \href{http://dx.doi.org/10.1103/PhysRevD.46.2553}{{\em Phys. Rev. D} {\bf 46}
  (1992)  2553--2559}, \href{http://arxiv.org/abs/0911.2064}{{\tt
  arXiv:0911.2064 [gr-qc]}}.

\bibitem{Folacci:1996dv}
A.~Folacci, ``{Toy model for the zero mode problem in the conformal sector of
  de Sitter quantum gravity},''
  \href{http://dx.doi.org/10.1103/PhysRevD.53.3108}{{\em Phys. Rev. D} {\bf 53}
  (1996)  3108--3117}.

\bibitem{Shaynkman:2000ts}
O.~V. Shaynkman and M.~A. Vasiliev, ``{Scalar field in any dimension from the
  higher spin gauge theory perspective},''
  \href{http://dx.doi.org/10.1007/BF02551402}{{\em Theor. Math. Phys.} {\bf
  123} (2000)  683--700}, \href{http://arxiv.org/abs/hep-th/0003123}{{\tt
  arXiv:hep-th/0003123}}.

\bibitem{Gazeau:2010mn}
J.-P. Gazeau, P.~Siegl, and A.~Youssef, ``{Krein Spaces in de Sitter Quantum
  Theories},'' \href{http://dx.doi.org/10.3842/SIGMA.2010.011}{{\em SIGMA} {\bf
  6} (2010)  011}, \href{http://arxiv.org/abs/1001.4810}{{\tt arXiv:1001.4810
  [hep-th]}}.

\bibitem{Bros:2010wa}
J.~Bros, H.~Epstein, and U.~Moschella, ``{Scalar tachyons in the de Sitter
  universe},'' \href{http://dx.doi.org/10.1007/s11005-010-0406-4}{{\em Lett.
  Math. Phys.} {\bf 93} (2010)  203--211},
  \href{http://arxiv.org/abs/1003.1396}{{\tt arXiv:1003.1396 [hep-th]}}.

\bibitem{Goon:2011qf}
G.~Goon, K.~Hinterbichler, and M.~Trodden, ``{Symmetries for Galileons and DBI
  scalars on curved space},''
  \href{http://dx.doi.org/10.1088/1475-7516/2011/07/017}{{\em JCAP} {\bf 07}
  (2011)  017}, \href{http://arxiv.org/abs/1103.5745}{{\tt arXiv:1103.5745
  [hep-th]}}.

\bibitem{Goon:2011uw}
G.~Goon, K.~Hinterbichler, and M.~Trodden, ``{A New Class of Effective Field
  Theories from Embedded Branes},''
  \href{http://dx.doi.org/10.1103/PhysRevLett.106.231102}{{\em Phys. Rev.
  Lett.} {\bf 106} (2011)  231102}, \href{http://arxiv.org/abs/1103.6029}{{\tt
  arXiv:1103.6029 [hep-th]}}.

\bibitem{Burrage:2011bt}
C.~Burrage, C.~de~Rham, and L.~Heisenberg, ``{de Sitter Galileon},''
  \href{http://dx.doi.org/10.1088/1475-7516/2011/05/025}{{\em JCAP} {\bf 05}
  (2011)  025}, \href{http://arxiv.org/abs/1104.0155}{{\tt arXiv:1104.0155
  [hep-th]}}.

\bibitem{Epstein:2014jaa}
H.~Epstein and U.~Moschella, ``{de Sitter tachyons and related topics},''
  \href{http://dx.doi.org/10.1007/s00220-015-2308-x}{{\em Commun. Math. Phys.}
  {\bf 336} (2015) no.~1, 381--430}, \href{http://arxiv.org/abs/1403.3319}{{\tt
  arXiv:1403.3319 [hep-th]}}.

\bibitem{Chekmenev:2015kzf}
A.~Chekmenev and M.~Grigoriev, ``{Boundary values of mixed-symmetry massless
  fields in AdS space},''
  \href{http://dx.doi.org/10.1016/j.nuclphysb.2016.10.006}{{\em Nucl. Phys. B}
  {\bf 913} (2016)  769--791}, \href{http://arxiv.org/abs/1512.06443}{{\tt
  arXiv:1512.06443 [hep-th]}}.

\bibitem{deBoer:2015kda}
J.~de~Boer, M.~P. Heller, R.~C. Myers, and Y.~Neiman, ``{Holographic de Sitter
  Geometry from Entanglement in Conformal Field Theory},''
  \href{http://dx.doi.org/10.1103/PhysRevLett.116.061602}{{\em Phys. Rev.
  Lett.} {\bf 116} (2016) no.~6, 061602},
  \href{http://arxiv.org/abs/1509.00113}{{\tt arXiv:1509.00113 [hep-th]}}.

\bibitem{deBoer:2016pqk}
J.~de~Boer, F.~M. Haehl, M.~P. Heller, and R.~C. Myers, ``{Entanglement,
  holography and causal diamonds},''
  \href{http://dx.doi.org/10.1007/JHEP08(2016)162}{{\em JHEP} {\bf 08} (2016)
  162}, \href{http://arxiv.org/abs/1606.03307}{{\tt arXiv:1606.03307
  [hep-th]}}.

\bibitem{Baumann:2017jvh}
D.~Baumann, G.~Goon, H.~Lee, and G.~L. Pimentel, ``{Partially Massless Fields
  During Inflation},'' \href{http://dx.doi.org/10.1007/JHEP04(2018)140}{{\em
  JHEP} {\bf 04} (2018)  140}, \href{http://arxiv.org/abs/1712.06624}{{\tt
  arXiv:1712.06624 [hep-th]}}.

\bibitem{Arkani-Hamed:2018kmz}
N.~Arkani-Hamed, D.~Baumann, H.~Lee, and G.~L. Pimentel, ``{The Cosmological
  Bootstrap: Inflationary Correlators from Symmetries and Singularities},''
  \href{http://dx.doi.org/10.1007/JHEP04(2020)105}{{\em JHEP} {\bf 04} (2020)
  105}, \href{http://arxiv.org/abs/1811.00024}{{\tt arXiv:1811.00024
  [hep-th]}}.

\bibitem{Bonifacio:2018zex}
J.~Bonifacio, K.~Hinterbichler, A.~Joyce, and R.~A. Rosen, ``{Shift Symmetries
  in (Anti) de Sitter Space},''
  \href{http://dx.doi.org/10.1007/JHEP02(2019)178}{{\em JHEP} {\bf 02} (2019)
  178}, \href{http://arxiv.org/abs/1812.08167}{{\tt arXiv:1812.08167
  [hep-th]}}.

\bibitem{Hinterbichler:2022vcc}
K.~Hinterbichler, ``{Shift symmetries for p-forms and mixed symmetry fields on
  (A)dS},'' \href{http://dx.doi.org/10.1007/JHEP11(2022)015}{{\em JHEP} {\bf
  11} (2022)  015}, \href{http://arxiv.org/abs/2207.03494}{{\tt
  arXiv:2207.03494 [hep-th]}}.

\bibitem{Volkov:1972jx}
D.~V. Volkov and V.~P. Akulov, ``{Possible universal neutrino interaction},''
  {\em JETP Lett.} {\bf 16} (1972)  438--440.

\bibitem{Volkov:1973ix}
D.~V. Volkov and V.~P. Akulov, ``{Is the Neutrino a Goldstone Particle?},''
  \href{http://dx.doi.org/10.1016/0370-2693(73)90490-5}{{\em Phys. Lett. B}
  {\bf 46} (1973)  109--110}.

\bibitem{Deser:1977uq}
S.~Deser and B.~Zumino, ``{Broken Supersymmetry and Supergravity},''
  \href{http://dx.doi.org/10.1103/PhysRevLett.38.1433}{{\em Phys. Rev. Lett.}
  {\bf 38} (1977)  1433--1436}.

\bibitem{Zumino:1977av}
B.~Zumino, ``{Nonlinear Realization of Supersymmetry in de Sitter Space},''
  \href{http://dx.doi.org/10.1016/0550-3213(77)90211-5}{{\em Nucl. Phys. B}
  {\bf 127} (1977)  189--201}.

\bibitem{Casalbuoni:1988sx}
R.~Casalbuoni, S.~De~Curtis, D.~Dominici, F.~Feruglio, and R.~Gatto, ``When
  does supergravity become strong?,''
  \href{http://dx.doi.org/10.1016/0370-2693(89)91123-4}{{\em Phys. Lett. B}
  {\bf 216} (1989)  325}. [Erratum: Phys.Lett.B 229, 439 (1989)].

\bibitem{10.1063/1.1665471}
F.~Schwarz, ``{Unitary Irreducible Representations of the Groups SO(n, 1)},''
  \href{http://dx.doi.org/10.1063/1.1665471}{{\em Journal of Mathematical
  Physics} {\bf 12} (1971) no.~1, 131--139}.

\bibitem{Bonifacio:2021mrf}
J.~Bonifacio, K.~Hinterbichler, A.~Joyce, and D.~Roest, ``{Exceptional scalar
  theories in de Sitter space},''
  \href{http://dx.doi.org/10.1007/JHEP04(2022)128}{{\em JHEP} {\bf 04} (2022)
  128}, \href{http://arxiv.org/abs/2112.12151}{{\tt arXiv:2112.12151
  [hep-th]}}.

\bibitem{DeRham:2018axr}
C.~De~Rham, K.~Hinterbichler, and L.~A. Johnson, ``{On the (A)dS Decoupling
  Limits of Massive Gravity},''
  \href{http://dx.doi.org/10.1007/JHEP09(2018)154}{{\em JHEP} {\bf 09} (2018)
  154}, \href{http://arxiv.org/abs/1807.08754}{{\tt arXiv:1807.08754
  [hep-th]}}.

\bibitem{Bonifacio:2019hrj}
J.~Bonifacio, K.~Hinterbichler, L.~A. Johnson, and A.~Joyce, ``{Shift-Symmetric
  Spin-1 Theories},'' \href{http://dx.doi.org/10.1007/JHEP09(2019)029}{{\em
  JHEP} {\bf 09} (2019)  029}, \href{http://arxiv.org/abs/1906.10692}{{\tt
  arXiv:1906.10692 [hep-th]}}.

\bibitem{Bogers:2018zeg}
M.~P. Bogers and T.~Brauner, ``{Lie-algebraic classification of effective
  theories with enhanced soft limits},''
  \href{http://dx.doi.org/10.1007/JHEP05(2018)076}{{\em JHEP} {\bf 05} (2018)
  076}, \href{http://arxiv.org/abs/1803.05359}{{\tt arXiv:1803.05359
  [hep-th]}}.

\bibitem{Roest:2019oiw}
D.~Roest, D.~Stefanyszyn, and P.~Werkman, ``{An Algebraic Classification of
  Exceptional EFTs},'' \href{http://dx.doi.org/10.1007/JHEP08(2019)081}{{\em
  JHEP} {\bf 08} (2019)  081}, \href{http://arxiv.org/abs/1903.08222}{{\tt
  arXiv:1903.08222 [hep-th]}}.

\bibitem{Khoury:2011da}
J.~Khoury, J.-L. Lehners, and B.~A. Ovrut, ``{Supersymmetric Galileons},''
  \href{http://dx.doi.org/10.1103/PhysRevD.84.043521}{{\em Phys. Rev. D} {\bf
  84} (2011)  043521}, \href{http://arxiv.org/abs/1103.0003}{{\tt
  arXiv:1103.0003 [hep-th]}}.

\bibitem{Koehn:2013hk}
M.~Koehn, J.-L. Lehners, and B.~Ovrut, ``{Supersymmetric cubic Galileons have
  ghosts},'' \href{http://dx.doi.org/10.1103/PhysRevD.88.023528}{{\em Phys.
  Rev. D} {\bf 88} (2013) no.~2, 023528},
  \href{http://arxiv.org/abs/1302.0840}{{\tt arXiv:1302.0840 [hep-th]}}.

\bibitem{Farakos:2013fne}
F.~Farakos, C.~Germani, and A.~Kehagias, ``{On ghost-free supersymmetric
  galileons},'' \href{http://dx.doi.org/10.1007/JHEP11(2013)045}{{\em JHEP}
  {\bf 11} (2013)  045}, \href{http://arxiv.org/abs/1306.2961}{{\tt
  arXiv:1306.2961 [hep-th]}}.

\bibitem{Kamimura:2013mia}
K.~Kamimura and S.~Onda, ``{Contractions of AdS brane algebra and superGalileon
  Lagrangians},'' \href{http://dx.doi.org/10.1063/1.4810765}{{\em J. Math.
  Phys.} {\bf 54} (2013)  062503}, \href{http://arxiv.org/abs/1303.5506}{{\tt
  arXiv:1303.5506 [hep-th]}}.

\bibitem{Queiruga:2016yzd}
J.~M. Queiruga, ``{Supersymmetric galileons and auxiliary fields in 2+1
  dimensions},'' \href{http://dx.doi.org/10.1103/PhysRevD.95.125001}{{\em Phys.
  Rev. D} {\bf 95} (2017) no.~12, 125001},
  \href{http://arxiv.org/abs/1612.04727}{{\tt arXiv:1612.04727 [hep-th]}}.

\bibitem{Elvang:2017mdq}
H.~Elvang, M.~Hadjiantonis, C.~R.~T. Jones, and S.~Paranjape, ``{On the
  Supersymmetrization of Galileon Theories in Four Dimensions},''
  \href{http://dx.doi.org/10.1016/j.physletb.2018.04.032}{{\em Phys. Lett. B}
  {\bf 781} (2018)  656--663}, \href{http://arxiv.org/abs/1712.09937}{{\tt
  arXiv:1712.09937 [hep-th]}}.

\bibitem{Roest:2017uga}
D.~Roest, P.~Werkman, and Y.~Yamada, ``{Internal Supersymmetry and Small-field
  Goldstini},'' \href{http://dx.doi.org/10.1007/JHEP05(2018)190}{{\em JHEP}
  {\bf 05} (2018)  190}, \href{http://arxiv.org/abs/1710.02480}{{\tt
  arXiv:1710.02480 [hep-th]}}.

\bibitem{Deen:2017dpm}
R.~Deen and B.~Ovrut, ``{$N =1$ supergravitational heterotic galileons},''
  \href{http://dx.doi.org/10.1007/JHEP11(2017)026}{{\em JHEP} {\bf 11} (2017)
  026}, \href{http://arxiv.org/abs/1707.05305}{{\tt arXiv:1707.05305
  [hep-th]}}.

\bibitem{Deen:2017jqv}
R.~Deen and B.~Ovrut, ``{Supergravitational Conformal Galileons},''
  \href{http://dx.doi.org/10.1007/JHEP08(2017)014}{{\em JHEP} {\bf 08} (2017)
  014}, \href{http://arxiv.org/abs/1705.06729}{{\tt arXiv:1705.06729
  [hep-th]}}.

\bibitem{Elvang:2018dco}
H.~Elvang, M.~Hadjiantonis, C.~R.~T. Jones, and S.~Paranjape, ``{Soft Bootstrap
  and Supersymmetry},'' \href{http://dx.doi.org/10.1007/JHEP01(2019)195}{{\em
  JHEP} {\bf 01} (2019)  195}, \href{http://arxiv.org/abs/1806.06079}{{\tt
  arXiv:1806.06079 [hep-th]}}.

\bibitem{Roest:2019dxy}
D.~Roest, D.~Stefanyszyn, and P.~Werkman, ``{An Algebraic Classification of
  Exceptional EFTs Part II: Supersymmetry},''
  \href{http://dx.doi.org/10.1007/JHEP11(2019)077}{{\em JHEP} {\bf 11} (2019)
  077}, \href{http://arxiv.org/abs/1905.05872}{{\tt arXiv:1905.05872
  [hep-th]}}.

\bibitem{Ivanov:1979ft}
E.~A. Ivanov and A.~S. Sorin, ``{{Wess-Zumino} Model as Linear Sigma Model of
  Spontaneously Broken Conformal and Osp(1,4) Supersymmetries},'' {\em Sov. J.
  Nucl. Phys.} {\bf 30} (1979)  440.

\bibitem{Garcia-Saenz:2018wnw}
S.~Garcia-Saenz, K.~Hinterbichler, and R.~A. Rosen, ``{Supersymmetric Partially
  Massless Fields and Non-Unitary Superconformal Representations},''
  \href{http://dx.doi.org/10.1007/JHEP11(2018)166}{{\em JHEP} {\bf 11} (2018)
  166}, \href{http://arxiv.org/abs/1810.01881}{{\tt arXiv:1810.01881
  [hep-th]}}.

\bibitem{Buchbinder:2019olk}
I.~L. Buchbinder, M.~V. Khabarov, T.~V. Snegirev, and Y.~M. Zinoviev,
  ``{Lagrangian description of the partially massless higher spin N = 1
  supermultiplets in AdS$_{4}$ space},''
  \href{http://dx.doi.org/10.1007/JHEP08(2019)116}{{\em JHEP} {\bf 08} (2019)
  116}, \href{http://arxiv.org/abs/1904.01959}{{\tt arXiv:1904.01959
  [hep-th]}}.

\bibitem{Bittermann:2020xkl}
N.~Bittermann, S.~Garcia-Saenz, K.~Hinterbichler, and R.~A. Rosen, ``{$
  \mathcal{N} $ = 2 supersymmetric partially massless fields and other exotic
  non-unitary superconformal representations},''
  \href{http://dx.doi.org/10.1007/JHEP08(2021)115}{{\em JHEP} {\bf 08} (2021)
  115}, \href{http://arxiv.org/abs/2011.05994}{{\tt arXiv:2011.05994
  [hep-th]}}.

\bibitem{Joung:2015jza}
E.~Joung and K.~Mkrtchyan, ``{Partially-massless higher-spin algebras and their
  finite-dimensional truncations},''
  \href{http://dx.doi.org/10.1007/JHEP01(2016)003}{{\em JHEP} {\bf 01} (2016)
  003}, \href{http://arxiv.org/abs/1508.07332}{{\tt arXiv:1508.07332
  [hep-th]}}.

\bibitem{Bekaert:2013zya}
X.~Bekaert and M.~Grigoriev, ``{Higher order singletons, partially massless
  fields and their boundary values in the ambient approach},''
  \href{http://dx.doi.org/10.1016/j.nuclphysb.2013.08.015}{{\em Nucl. Phys. B}
  {\bf 876} (2013)  667--714}, \href{http://arxiv.org/abs/1305.0162}{{\tt
  arXiv:1305.0162 [hep-th]}}.

\bibitem{Basile:2014wua}
T.~Basile, X.~Bekaert, and N.~Boulanger, ``{Flato-Fronsdal theorem for
  higher-order singletons},''
  \href{http://dx.doi.org/10.1007/JHEP11(2014)131}{{\em JHEP} {\bf 11} (2014)
  131}, \href{http://arxiv.org/abs/1410.7668}{{\tt arXiv:1410.7668 [hep-th]}}.

\bibitem{Alkalaev:2014nsa}
K.~B. Alkalaev, M.~Grigoriev, and E.~D. Skvortsov, ``{Uniformizing higher-spin
  equations},'' \href{http://dx.doi.org/10.1088/1751-8113/48/1/015401}{{\em J.
  Phys. A} {\bf 48} (2015) no.~1, 015401},
  \href{http://arxiv.org/abs/1409.6507}{{\tt arXiv:1409.6507 [hep-th]}}.

\bibitem{Brust:2016zns}
C.~Brust and K.~Hinterbichler, ``{Partially Massless Higher-Spin Theory},''
  \href{http://dx.doi.org/10.1007/JHEP02(2017)086}{{\em JHEP} {\bf 02} (2017)
  086}, \href{http://arxiv.org/abs/1610.08510}{{\tt arXiv:1610.08510
  [hep-th]}}.

\bibitem{Binder:2020raz}
D.~J. Binder, D.~Z. Freedman, and S.~S. Pufu, ``{A bispinor formalism for
  spinning Witten diagrams},''
  \href{http://dx.doi.org/10.1007/JHEP02(2022)040}{{\em JHEP} {\bf 02} (2022)
  040}, \href{http://arxiv.org/abs/2003.07448}{{\tt arXiv:2003.07448
  [hep-th]}}.

\bibitem{Deser:2001pe}
S.~Deser and A.~Waldron, ``{Gauge invariances and phases of massive higher
  spins in (A)dS},''
  \href{http://dx.doi.org/10.1103/PhysRevLett.87.031601}{{\em Phys. Rev. Lett.}
  {\bf 87} (2001)  031601}, \href{http://arxiv.org/abs/hep-th/0102166}{{\tt
  arXiv:hep-th/0102166}}.

\bibitem{Deser:2001xr}
S.~Deser and A.~Waldron, ``{Null propagation of partially massless higher spins
  in (A)dS and cosmological constant speculations},''
  \href{http://dx.doi.org/10.1016/S0370-2693(01)00756-0}{{\em Phys. Lett. B}
  {\bf 513} (2001)  137--141}, \href{http://arxiv.org/abs/hep-th/0105181}{{\tt
  arXiv:hep-th/0105181}}.

\bibitem{Deser:2003gw}
S.~Deser and A.~Waldron, ``{Arbitrary spin representations in de Sitter from dS
  / CFT with applications to dS supergravity},''
  \href{http://dx.doi.org/10.1016/S0550-3213(03)00348-1}{{\em Nucl. Phys. B}
  {\bf 662} (2003)  379--392}, \href{http://arxiv.org/abs/hep-th/0301068}{{\tt
  arXiv:hep-th/0301068}}.

\bibitem{Fang:1979hq}
J.~Fang and C.~Fronsdal, ``{Massless, Half Integer Spin Fields in De Sitter
  Space},'' \href{http://dx.doi.org/10.1103/PhysRevD.22.1361}{{\em Phys. Rev.
  D} {\bf 22} (1980)  1361}.

\bibitem{Aragone:1980rk}
C.~Aragone and S.~Deser, ``{Higher Spin Vierbein Gauge Fermions and
  Hypergravities},'' \href{http://dx.doi.org/10.1016/0550-3213(80)90153-4}{{\em
  Nucl. Phys. B} {\bf 170} (1980)  329--352}.

\bibitem{Vasiliev:1987tk}
M.~A. Vasiliev, ``{Free Massless Fermionic Fields of Arbitrary Spin in
  $d$-dimensional De Sitter Space},''
  \href{http://dx.doi.org/10.1016/0550-3213(88)90161-7}{{\em Nucl. Phys. B}
  {\bf 301} (1988)  26--68}.

\bibitem{Basile:2016aen}
T.~Basile, X.~Bekaert, and N.~Boulanger, ``{Mixed-symmetry fields in de Sitter
  space: a group theoretical glance},''
  \href{http://dx.doi.org/10.1007/JHEP05(2017)081}{{\em JHEP} {\bf 05} (2017)
  081}, \href{http://arxiv.org/abs/1612.08166}{{\tt arXiv:1612.08166
  [hep-th]}}.

\bibitem{Letsios:2023qzq}
V.~A. Letsios, ``{(Non-)unitarity of strictly and partially massless fermions
  on de Sitter space},'' \href{http://arxiv.org/abs/2303.00420}{{\tt
  arXiv:2303.00420 [hep-th]}}.

\bibitem{Campoleoni:2017vds}
A.~Campoleoni, M.~Henneaux, S.~H\"ortner, and A.~Leonard, ``{Higher-spin
  charges in Hamiltonian form. II. Fermi fields},''
  \href{http://dx.doi.org/10.1007/JHEP02(2017)058}{{\em JHEP} {\bf 02} (2017)
  058}, \href{http://arxiv.org/abs/1701.05526}{{\tt arXiv:1701.05526
  [hep-th]}}.

\bibitem{Metsaev:1998xg}
R.~R. Metsaev, ``{Fermionic fields in the d-dimensional anti-de Sitter
  space-time},'' \href{http://dx.doi.org/10.1016/S0370-2693(97)01446-9}{{\em
  Phys. Lett. B} {\bf 419} (1998)  49--56},
  \href{http://arxiv.org/abs/hep-th/9802097}{{\tt arXiv:hep-th/9802097}}.

\bibitem{Zinoviev:2009vy}
Y.~M. Zinoviev, ``{Frame-like gauge invariant formulation for mixed symmetry
  fermionic fields},''
  \href{http://dx.doi.org/10.1016/j.nuclphysb.2009.06.008}{{\em Nucl. Phys. B}
  {\bf 821} (2009)  21--47}, \href{http://arxiv.org/abs/0904.0549}{{\tt
  arXiv:0904.0549 [hep-th]}}.

\bibitem{Skvortsov:2010nh}
E.~D. Skvortsov and Y.~M. Zinoviev, ``{Frame-like Actions for Massless
  Mixed-Symmetry Fields in Minkowski space. Fermions},''
  \href{http://dx.doi.org/10.1016/j.nuclphysb.2010.10.012}{{\em Nucl. Phys. B}
  {\bf 843} (2011)  559--569}, \href{http://arxiv.org/abs/1007.4944}{{\tt
  arXiv:1007.4944 [hep-th]}}.

\bibitem{Reshetnyak:2012ec}
A.~A. Reshetnyak, ``{General Lagrangian Formulation for Higher Spin Fields with
  Arbitrary Index Symmetry. 2. Fermionic fields},''
  \href{http://dx.doi.org/10.1016/j.nuclphysb.2012.12.010}{{\em Nucl. Phys. B}
  {\bf 869} (2013)  523--597}, \href{http://arxiv.org/abs/1211.1273}{{\tt
  arXiv:1211.1273 [hep-th]}}.

\bibitem{Metsaev:1995re}
R.~R. Metsaev, ``{Massless mixed symmetry bosonic free fields in d-dimensional
  anti-de Sitter space-time},''
  \href{http://dx.doi.org/10.1016/0370-2693(95)00563-Z}{{\em Phys. Lett. B}
  {\bf 354} (1995)  78--84}.

\bibitem{Metsaev:1997nj}
R.~R. Metsaev, ``{Arbitrary spin massless bosonic fields in d-dimensional
  anti-de Sitter space},'' \href{http://dx.doi.org/10.1007/BFb0104614}{{\em
  Lect. Notes Phys.} {\bf 524} (1999)  331--340},
  \href{http://arxiv.org/abs/hep-th/9810231}{{\tt arXiv:hep-th/9810231}}.

\bibitem{Alkalaev:2003qv}
K.~B. Alkalaev, O.~V. Shaynkman, and M.~A. Vasiliev, ``{On the frame - like
  formulation of mixed symmetry massless fields in (A)dS(d)},''
  \href{http://dx.doi.org/10.1016/j.nuclphysb.2004.05.031}{{\em Nucl. Phys. B}
  {\bf 692} (2004)  363--393}, \href{http://arxiv.org/abs/hep-th/0311164}{{\tt
  arXiv:hep-th/0311164}}.

\bibitem{Boulanger:2008up}
N.~Boulanger, C.~Iazeolla, and P.~Sundell, ``{Unfolding Mixed-Symmetry Fields
  in AdS and the BMV Conjecture: I. General Formalism},''
  \href{http://dx.doi.org/10.1088/1126-6708/2009/07/013}{{\em JHEP} {\bf 07}
  (2009)  013}, \href{http://arxiv.org/abs/0812.3615}{{\tt arXiv:0812.3615
  [hep-th]}}.

\bibitem{Boulanger:2008kw}
N.~Boulanger, C.~Iazeolla, and P.~Sundell, ``{Unfolding Mixed-Symmetry Fields
  in AdS and the BMV Conjecture. II. Oscillator Realization},''
  \href{http://dx.doi.org/10.1088/1126-6708/2009/07/014}{{\em JHEP} {\bf 07}
  (2009)  014}, \href{http://arxiv.org/abs/0812.4438}{{\tt arXiv:0812.4438
  [hep-th]}}.

\bibitem{Skvortsov:2009zu}
E.~D. Skvortsov, ``{Gauge fields in (A)dS(d) and Connections of its symmetry
  algebra},'' \href{http://dx.doi.org/10.1088/1751-8113/42/38/385401}{{\em J.
  Phys. A} {\bf 42} (2009)  385401}, \href{http://arxiv.org/abs/0904.2919}{{\tt
  arXiv:0904.2919 [hep-th]}}.

\bibitem{Skvortsov:2009nv}
E.~D. Skvortsov, ``{Gauge fields in (A)dS(d) within the unfolded approach:
  algebraic aspects},'' \href{http://dx.doi.org/10.1007/JHEP01(2010)106}{{\em
  JHEP} {\bf 01} (2010)  106}, \href{http://arxiv.org/abs/0910.3334}{{\tt
  arXiv:0910.3334 [hep-th]}}.

\bibitem{Brink:2000ag}
L.~Brink, R.~R. Metsaev, and M.~A. Vasiliev, ``{How massless are massless
  fields in AdS(d)},''
  \href{http://dx.doi.org/10.1016/S0550-3213(00)00402-8}{{\em Nucl. Phys. B}
  {\bf 586} (2000)  183--205}, \href{http://arxiv.org/abs/hep-th/0005136}{{\tt
  arXiv:hep-th/0005136}}.

\bibitem{Costa:2014rya}
M.~S. Costa and T.~Hansen, ``{Conformal correlators of mixed-symmetry
  tensors},'' \href{http://dx.doi.org/10.1007/JHEP02(2015)151}{{\em JHEP} {\bf
  02} (2015)  151}, \href{http://arxiv.org/abs/1411.7351}{{\tt arXiv:1411.7351
  [hep-th]}}.

\bibitem{Metsaev:1997hi}
R.~R. Metsaev, ``{Free totally (anti)symmetric massless fermionic fields in
  d-dimensional anti-de Sitter space},''
  \href{http://dx.doi.org/10.1088/0264-9381/14/5/008}{{\em Class. Quant. Grav.}
  {\bf 14} (1997)  L115--L121}, \href{http://arxiv.org/abs/hep-th/9707066}{{\tt
  arXiv:hep-th/9707066}}.

\bibitem{Buchbinder:2009pa}
I.~L. Buchbinder, V.~A. Krykhtin, and L.~L. Ryskina, ``{Lagrangian formulation
  of massive fermionic totally antisymmetric tensor field theory in AdS(d)
  space},'' \href{http://dx.doi.org/10.1016/j.nuclphysb.2009.04.014}{{\em Nucl.
  Phys. B} {\bf 819} (2009)  453--477},
  \href{http://arxiv.org/abs/0902.1471}{{\tt arXiv:0902.1471 [hep-th]}}.

\bibitem{Zinoviev:2009wh}
Y.~M. Zinoviev, ``{Note on antisymmetric spin-tensors},''
  \href{http://dx.doi.org/10.1088/1126-6708/2009/04/035}{{\em JHEP} {\bf 04}
  (2009)  035}, \href{http://arxiv.org/abs/0903.0262}{{\tt arXiv:0903.0262
  [hep-th]}}.

\bibitem{Campoleoni:2009gs}
A.~Campoleoni, D.~Francia, J.~Mourad, and A.~Sagnotti, ``{Unconstrained Higher
  Spins of Mixed Symmetry. II. Fermi Fields},''
  \href{http://dx.doi.org/10.1016/j.nuclphysb.2009.08.025}{{\em Nucl. Phys. B}
  {\bf 828} (2010)  405--514}, \href{http://arxiv.org/abs/0904.4447}{{\tt
  arXiv:0904.4447 [hep-th]}}.

\bibitem{Lekeu:2021oti}
V.~Lekeu and Y.~Zhang, ``{On the quantisation and anomalies of antisymmetric
  tensor-spinors},'' \href{http://dx.doi.org/10.1007/JHEP11(2021)078}{{\em
  JHEP} {\bf 11} (2021)  078}, \href{http://arxiv.org/abs/2109.03963}{{\tt
  arXiv:2109.03963 [hep-th]}}.

\bibitem{Wang:2023iqt}
Y.-N. Wang and Y.~Zhang, ``{Fermionic Higher-form Symmetries},''
  \href{http://arxiv.org/abs/2303.12633}{{\tt arXiv:2303.12633 [hep-th]}}.

\bibitem{Pethybridge:2021rwf}
B.~Pethybridge and V.~Schaub, ``{Tensors and spinors in de Sitter space},''
  \href{http://dx.doi.org/10.1007/JHEP06(2022)123}{{\em JHEP} {\bf 06} (2022)
  123}, \href{http://arxiv.org/abs/2111.14899}{{\tt arXiv:2111.14899
  [hep-th]}}.

\bibitem{Dirac-ambient}
P.~A.~M. Dirac, ``Wave equations in conformal space,''
  \href{http://dx.doi.org/https://doi.org/10.2307/1968455}{{\em Annals of
  Mathematics} {\bf 37} (1936) no.~2, 429--442}.

\bibitem{Bekaert:2010hk}
X.~Bekaert and E.~Meunier, ``{Higher spin interactions with scalar matter on
  constant curvature spacetimes: conserved current and cubic coupling
  generating functions},''
  \href{http://dx.doi.org/10.1007/JHEP11(2010)116}{{\em JHEP} {\bf 11} (2010)
  116}, \href{http://arxiv.org/abs/1007.4384}{{\tt arXiv:1007.4384 [hep-th]}}.

\bibitem{Takook:2014paa}
M.~V. Takook, ``{Quantum Field Theory in de Sitter Universe: Ambient Space
  Formalism},'' \href{http://arxiv.org/abs/1403.1204}{{\tt arXiv:1403.1204
  [gr-qc]}}.

\bibitem{Iliesiu:2015qra}
L.~Iliesiu, F.~Kos, D.~Poland, S.~S. Pufu, D.~Simmons-Duffin, and R.~Yacoby,
  ``{Bootstrapping 3D Fermions},''
  \href{http://dx.doi.org/10.1007/JHEP03(2016)120}{{\em JHEP} {\bf 03} (2016)
  120}, \href{http://arxiv.org/abs/1508.00012}{{\tt arXiv:1508.00012
  [hep-th]}}.

\bibitem{Nishida:2018opl}
M.~Nishida and K.~Tamaoka, ``{Fermions in Geodesic Witten Diagrams},''
  \href{http://dx.doi.org/10.1007/JHEP07(2018)149}{{\em JHEP} {\bf 07} (2018)
  149}, \href{http://arxiv.org/abs/1805.00217}{{\tt arXiv:1805.00217
  [hep-th]}}.

\bibitem{Schaub:2023scu}
V.~Schaub, ``{Spinors in (Anti-)de Sitter Space},''
  \href{http://arxiv.org/abs/2302.08535}{{\tt arXiv:2302.08535 [hep-th]}}.

\bibitem{Kac1977}
V.~G. Kac, ``{A sketch of Lie superalgebra theory},'' {\em Communications in
  Mathematical Physics} {\bf 53} (1977) no.~1, 31 -- 64.

\bibitem{Love:2006hv}
S.~T. Love, ``{Dynamics of (SUSY) AdS Space Isometry Breaking},''
  \href{http://dx.doi.org/10.1088/1751-8113/40/25/S60}{{\em J. Phys. A} {\bf
  40} (2007)  7049--7054}, \href{http://arxiv.org/abs/hep-th/0611199}{{\tt
  arXiv:hep-th/0611199}}.

\bibitem{Clark:2005ht}
T.~E. Clark, S.~T. Love, M.~Nitta, and T.~ter Veldhuis, ``{AdS(d+1)
  ---\ensuremath{>} AdS(d)},'' \href{http://dx.doi.org/10.1063/1.2048307}{{\em
  J. Math. Phys.} {\bf 46} (2005)  102304},
  \href{http://arxiv.org/abs/hep-th/0501241}{{\tt arXiv:hep-th/0501241}}.

\bibitem{Hinterbichler:2012fr}
K.~Hinterbichler, A.~Joyce, J.~Khoury, and G.~E.~J. Miller, ``{DBI Realizations
  of the Pseudo-Conformal Universe and Galilean Genesis Scenarios},''
  \href{http://dx.doi.org/10.1088/1475-7516/2012/12/030}{{\em JCAP} {\bf 12}
  (2012)  030}, \href{http://arxiv.org/abs/1209.5742}{{\tt arXiv:1209.5742
  [hep-th]}}.

\bibitem{Clark:2005ma}
T.~E. Clark and S.~T. Love, ``{Nonlinear realization of supersymmetric AdS
  space isometries},'' \href{http://dx.doi.org/10.1103/PhysRevD.73.025001}{{\em
  Phys. Rev. D} {\bf 73} (2006)  025001},
  \href{http://arxiv.org/abs/hep-th/0510274}{{\tt arXiv:hep-th/0510274}}.

\bibitem{Delduc:2001tb}
F.~Delduc, E.~Ivanov, and S.~Krivonos, ``{Partial supersymmetry breaking and
  AdS(4) supermembrane},''
  \href{http://dx.doi.org/10.1016/S0370-2693(02)01260-1}{{\em Phys. Lett. B}
  {\bf 529} (2002)  233--240}, \href{http://arxiv.org/abs/hep-th/0111106}{{\tt
  arXiv:hep-th/0111106}}.

\bibitem{Heidenreich:1982rz}
W.~Heidenreich, ``All linear unitary irreducible representations of de sitter
  supersymmetry with positive energy,''
  \href{http://dx.doi.org/10.1016/0370-2693(82)91038-3}{{\em Phys. Lett. B}
  {\bf 110} (1982)  461--464}.

\bibitem{Coleman:1969sm}
S.~R. Coleman, J.~Wess, and B.~Zumino, ``{Structure of phenomenological
  Lagrangians. 1.},'' \href{http://dx.doi.org/10.1103/PhysRev.177.2239}{{\em
  Phys. Rev.} {\bf 177} (1969)  2239--2247}.

\bibitem{Callan:1969sn}
C.~G. Callan, Jr., S.~R. Coleman, J.~Wess, and B.~Zumino, ``{Structure of
  phenomenological Lagrangians. 2.},''
  \href{http://dx.doi.org/10.1103/PhysRev.177.2247}{{\em Phys. Rev.} {\bf 177}
  (1969)  2247--2250}.

\bibitem{Zumino:1970tu}
B.~Zumino, ``{Effective Lagrangians and Broken Symmetries},'' in {\em Lectures
  on Elementary Particles and Quantum Field Theory}, vol.~2, p.~437.
\newblock Jan., 1970.

\bibitem{Volkov:1973vd}
D.~V. Volkov, ``{Phenomenological Lagrangians},'' {\em Fiz. Elem. Chast. Atom.
  Yadra} {\bf 4} (1973)  3--41.

\bibitem{Goon:2012dy}
G.~Goon, K.~Hinterbichler, A.~Joyce, and M.~Trodden, ``{Galileons as
  Wess-Zumino Terms},'' \href{http://dx.doi.org/10.1007/JHEP06(2012)004}{{\em
  JHEP} {\bf 06} (2012)  004}, \href{http://arxiv.org/abs/1203.3191}{{\tt
  arXiv:1203.3191 [hep-th]}}.

\end{thebibliography}\endgroup

\end{document}